\documentclass[11pt,twocolumn,english,american,journal]{IEEEtran}
\usepackage[T1]{fontenc}
\usepackage[latin9]{inputenc}
\usepackage{geometry}
\usepackage{textcomp}
\usepackage{amsmath}
\usepackage{amssymb}
\usepackage{graphicx}

\makeatletter
\let\oldforeign@language\foreign@language
\DeclareRobustCommand{\foreign@language}[1]{%
  \lowercase{\oldforeign@language{#1}}}

\usepackage[caption=false,font=footnotesize]{subfig}

\usepackage{colortbl}
\usepackage{cite}

\IEEEoverridecommandlockouts
\IEEEpubid{\makebox[\columnwidth]{978-1-5386-5541-2/18/\$31.00~
\copyright2018 IEEE \hfill} \hspace{\columnsep}
\makebox[\columnwidth]{ }}

\makeatother

\usepackage{babel}
\begin{document}
\title{The Generalized Droop Formula}
\author{\selectlanguage{english}%
Alberto Bononi,~\IEEEmembership{Senior Member,~IEEE,} Jean-Christophe
Antona, Matteo Lonardi, \emph{Student Member,~IEEE,} Alexis Carbo-Méseguer
and Paolo Serena,\emph{~Member,~IEEE }\thanks{Manuscript received xxxxxxxxx xx, xxxx; accepted xxxxxxxxx xx, xxxx.
Date of publication xxxxxxxxx xx, xxxx; date of current version xxxxxxxxx
xx, xxxx. }\thanks{A. Bononi, P. Serena, M. Lonardi are with the Dipartimento di Ingegneria
e Architettura, Universit\`a di Parma, Parma 43124, Italy (corresponding
author e-mail: alberto.bononi@unipr.it). J.-C. Antona and A. Carbo-Méseguer
are with Alcatel Submarine Networks, Villarceaux, France.}\thanks{Color versions of one or more of the figures in this paper are available
online at http://ieeexplore.ieee.org.}\thanks{Digital Object Identifier xx.xxxx/JLT.xxxx.xxxxxxx.}}
\markboth{\foreignlanguage{english}{Submitted to the IEEE Journal of Lightwave Technology, Aug. 27, 2019}}{Alberto Bononi\MakeLowercase{\emph{et al.}}}

\maketitle
\IEEEpubidadjcol
\selectlanguage{american}%
\begin{abstract}
We present a theoretical model that fully supports the recently disclosed
generalized droop formula (GDF) for calculating the signal-to-noise
ratio (SNR) of constant-output power (COP) amplified dispersion-uncompensated
coherent links operated at very low SNR. We compare the GDF to the
better known Gaussian noise (GN) model. \foreignlanguage{english}{A
key finding is that the end-to-end model underlying the GDF is a concatenation
of per-span first-order regular perturbation (RP1) models with end-span
power renormalization. This fact allows the GDF to well reproduce
the SNR of highly nonlinear systems, well beyond the RP1 limit underlying
the GN model. The GDF is successfully extended to the case where the
bandwidth/modes of the COP amplifiers are not entirely filled by the
transmitted multiplex. Finally, the GDF  is extended to constant-gain
(CG) amplifier chains and is shown to improve on known GN models of
highly nonlinear propagation with CG amplifiers.}
\end{abstract}

\selectlanguage{english}%

\selectlanguage{american}%

\section{Introduction\foreignlanguage{english}{\label{sec:Introduction}}}

\selectlanguage{english}%
\IEEEPARstart{\foreignlanguage{american}{A}}{}mplified spontaneous
emission (ASE)\foreignlanguage{american}{-induced signal droop in
constant output power (COP) amplifier chains was studied long ago
\cite{giles}. Today's submarine systems basically all use COP amplifiers,
but most analytical models for single-mode transmission do assume
constant-gain (CG) amplifiers, with a few exceptions (e.g., \cite{amir_droop}).
In the context of space-division multiplexed (SDM) submarine transmissions,
the term \textquotedblright droop\textquotedblright{} was introduced
a few years ago by Sinkin \emph{et al}. \cite{TEsubcom_ptl17,TEsubcom_jlt18},
who revived the droop problem in COP amplified SDM links. Antona \emph{et
al.} \cite{ASN_droop_ofc19,ASN_suboptic} recently proposed a new
expression of the received signal to noise ratio (SNR) in very-long
haul, low-SNR dispersion-uncompensated submarine links with coherent
detection, which we here call the generalized droop formula (GDF),
and showed that for single-mode COP-amplified links the GDF is more
precise than the SNR predicted by the widely-used Gaussian Noise (GN)
model \cite{poggio_ptl,grellier,GN_carena,GN_poggio}, which assumes
CG amplifiers. The GDF accounts for both signal and ASE droop, both
in the linear and in the nonlinear regime. Extensions of the GDF to
account for other sources of power rearrangement in the transmission
fiber were also proposed \cite{ASN_droop_ofc19,ASN_suboptic}.}

\selectlanguage{american}%
This paper, which is an extension of \cite{noiECOC20_subm}, aims
to recast the heuristic derivations in \cite{ASN_droop_ofc19,ASN_suboptic}
on solid theoretical footings, and to extend the comparisons of the
GDF against the GN model and its extension to COP amplifiers \cite{amir_droop}.
\foreignlanguage{english}{One of the key theoretical findings of this
paper is that the end-to-end model underlying the GDF is a concatenation
of per-span first-order regular perturbation (RP1) models \cite{noi_RP}
with end-span power renormalization. This fact allows the GDF to well
reproduce the SNR of highly nonlinear systems, well beyond the RP1
limit underlying the GN model \cite{poggio_arxiv,johannisson_GN,RP1_GN_arxiv}.
This multi-stage RP1 is reminiscent of multi-stage backpropagation
\cite{secondini,kumar} that combines the benefits of split-step and
perturbation-based approaches.}

The GDF theory is then extended to the more general case where significant
out-of-band ASE is present in the system, yielding a new SNR expression
that we call the COP-GDF. While working out the extended theory, we
find deep connections also with constant-gain (CG) amplifier chains
with significant nonlinear signal-ASE interactions, for which extensions
of the GN theory are known \cite{poggio_depletion-2,serena_S-NLI_interaction}.
We here propose a new formula, which we call the CG-GDF. All formulas
are checked against accurate split-step Fourier method (SSFM) simulations.
In particular, COP-GDF and CG-GDF always show the best match with
simulations among all known formulas.

The paper is organized as follows. Sections \ref{sec:Droop-induced-by}
and \ref{sec:Droop-induced-by-1} introduce the system model and derive
the additive and rearrangement droops. Sec. \ref{sec:SNR} derives
the basic GDF, discusses its implications and introduces upper and
lower bounds to the GDF. Sec. \ref{sec:Numerical-checks} presents
numerical comparisons of theory against simulations, and GDF against
both its approximations and against the GN formula. Sec. \ref{sec:The-failure-of}
extends the GDF to the case when ASE has a larger spectral occupancy
than the useful signal. Sec. \ref{sec:The-constant-gain-case} discusses
the CG case and derives the new CG-GDF expression. Sec. \ref{sec:Conclusions}
concludes the paper.

\section{\label{sec:Droop-induced-by} Droop induced by power addition}

Consider the transmission of a mode/wavelength division multiplexed
(M/WDM) signal, composed of $M$ spatial modes (each corresponding
to two orthogonal polarizations) each composed of an $N_{c}$-channel
Nyquist-WDM system \cite{nyq-WDM} over a total bandwidth $B$, along
a chain of $N$ \emph{identical} multi-mode/core fiber (MF) spans.
All spans have loss $\mathcal{L}<1$ and are followed by an end-span
amplifier having a total constant output power (COP) equal to $P$.
We assume the multiplex total launched power is $P$, and that loss
$\mathcal{L}$ and amplifier gain $G$ are the same at all wavelengths
and modes. We also assume the amplifier has a filter that suppresses
all out-of band/mode ASE noise, so that ASE and signal spectra are
flat over the same bandwidth $B$. The Nyquist-WDM assumption and
the assumption that ASE and signal exist on the same spectral range
will be relaxed in Section \ref{sec:The-failure-of}.

\selectlanguage{english}%
\begin{figure}
\centering

\includegraphics[width=1\columnwidth]{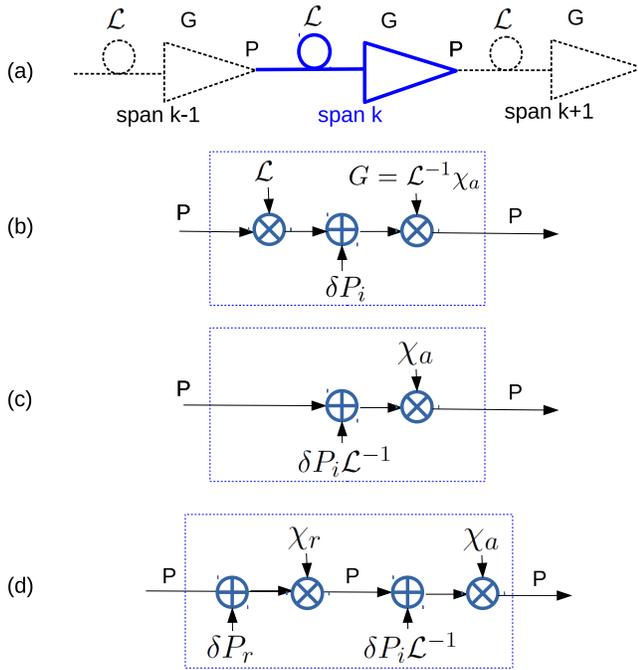}

\caption{\label{fig:block} \textbf{(a)} Chain of spans with loss $\mathcal{L}$
with amplifiers with constant output power $P$ and ASE equivalent
input power $\delta P_{i}$. Gain must be $G=\mathcal{L}^{-1}\chi_{a}$
(see text), with $\chi_{a}<1$ the ASE-induced power droop. \textbf{(b)}
Span power block diagram. \textbf{(c)} Equivalent diagram with loss
factored out. \textbf{(d) }Added block diagram also of fiber span
with redistribution power $\delta P_{r}$ and renormalization to $P$
by a redistribution power droop $\chi_{r}<1$.}
\end{figure}

\selectlanguage{american}%
As seen in Fig. \foreignlanguage{english}{\ref{fig:block}}(a), the
chain has at each span $k$ a total input power $P$, and a total
output power $P$. Hence in the ideal case of noiseless amplifiers,
the amplifier gain $G=\mathcal{L}^{-1}$ exactly compensates the loss.
The span block diagram for real amplifiers is shown in Fig. \foreignlanguage{english}{\ref{fig:block}}(b),
where an equivalent input ASE noise power $\delta P_{i}=Mh\nu FB$
(where $h$ is Planck's constant, $\nu$ is the multiplex center frequency,
$F$ is the noise figure \cite{Haus}, $B$ the amplification bandwidth
and $M$ is the number of modes) is injected in the amplifier, hence
the gain $G=\mathcal{L}^{-1}\chi_{a}$ must decrease by a \emph{droop}
factor $\chi_{a}<1$ from the ideal case to ``squeeze'' the transiting
signal and make room for the local ASE noise in the output power budget
$P$. By shifting back the term $\mathcal{L}^{-1}$ upstream of the
addition block (i.e., by ``factoring out'' the loss) the block of
Fig. \foreignlanguage{english}{\ref{fig:block}}(b) is seen to be
equivalent to that in Fig. \foreignlanguage{english}{\ref{fig:block}}(c),
from which we read the span input-output power budget as: $(P+\delta P_{i}\mathcal{L}^{-1})\chi_{a}=P$,
and thus deduce the droop 
\begin{equation}
\chi_{a}=(1+\frac{\delta P_{i}\mathcal{L}^{-1}}{P})^{-1}\triangleq(1+SNR_{a1}^{-1})^{-1}\label{eq:chia}
\end{equation}
where $\delta P_{i}\mathcal{L}^{-1}$ is the output ASE that would
be generated by an end-span amplifier with gain $\mathcal{L}^{-1}$,
and we implicitly defined SNR degraded at the single amplifier as
\begin{equation}
SNR_{a1}\triangleq\frac{P}{\delta P_{i}\mathcal{L}^{-1}}.\label{eq:SNRa1}
\end{equation}

The droop is in fact the total power gain (in fact, a loss) of each
amplified span, so that the desired multiplex signal power at the
output of the $N$-th amplifier is
\begin{equation}
P_{s}(N)=P\prod_{k=1}^{N}\mathcal{L}G=P\chi_{a}^{N}\label{eq:PsN}
\end{equation}
which tells us that the desired signal becomes weaker along the nominally
transparent line because of the accumulation of ASE which reduces
the amplifier gain $G$ because of the COP constraint. The accumulated
ASE at the output of the N-span chain (over all modes and amplified
WDM bandwidth $B$) is thus
\begin{equation}
P_{a}(N)=P-P_{s}(N)=P(1-\chi_{a}^{N}).\label{eq:PaN}
\end{equation}
\foreignlanguage{english}{ By equating (\ref{eq:PsN}),(\ref{eq:PaN})
we find that signal and ASE powers become equal at $N\cong\ln2\cdot SNR_{a1}$.}

Note that the above analysis remains unchanged if the amplifiers were
noiseless, but an external lumped crosstalk (e.g., power leaking from
a competing optical multiplex of power $P$ crossing an optical multiplexer/demultiplexer
together with our multiplex of interest at an optical node before
the final optical amplification, or, e.g., transmitter impairments
at the booster amplifier) of power $\delta P_{i}=\alpha_{ex}P$ were
injected in its place, where $\alpha_{ex}$ is the external crosstalk
coefficient. In presence of both ASE and external crosstalk the droop
$\chi_{a}$ in (\ref{eq:chia}), that we more generally call the \emph{addition
droop,} uses an added power $\delta P_{i}=Mh\nu FB+\alpha_{ex}P$,
where uncorrelation between the two noise sources is assumed when
summing power.

\section{\label{sec:Droop-induced-by-1}Droop induced by power redistribution}

The transmission fiber is indeed not ideal and operates a power redistribution
during propagation because of several physical mechanisms. Let's for
the moment concentrate on one of these, namely, the nonlinear Kerr
effect. Focus on Fig. 1(c) where the fiber loss has been factored
out. We now apply a first-order regular perturbation approximation
of the Kerr distortion generated within span $k$, called nonlinear
interference (NLI), as in the GN and similar perturbative models \cite{poggio_ptl,GN_poggio,grellier}.
We then impose that the power in/out of the fiber be conserved. Thus,
we get a power-flow diagram of the fiber+amplifier block as depicted
in Fig. 1(d), where now the power redistribution during propagation
appears as a new input sub-block in which a perturbation $\delta P_{r}=\alpha_{NL}P^{3}$
is added to the input signal $P$ ($\alpha_{NL}$ is the per-span
NLI coefficient \cite{Vacondio}), and then a \emph{redistribution
droop} $\chi_{r}$ forces the perturbed signal back to power $P$,
namely, $(P+\delta P_{r})\chi_{r}=P$. This yields
\begin{equation}
\chi_{r}=(1+\frac{\delta P_{r}}{P})^{-1}\triangleq(1+SNR_{r1}^{-1})^{-1}\label{eq:chir}
\end{equation}
where we implicitly defined the SNR degraded at the single amplifier
by the redistribution mechanism as 
\begin{equation}
SNR_{r1}\triangleq\frac{P}{\delta P_{r}}=\frac{1}{\alpha_{NL}P^{2}}\label{eq:SNRr1}
\end{equation}
where the second equality holds specifically for the NLI redistribution
mechanism.

In other terms, we first apply a per-span RP1 perturbation, and then
re-normalize signal plus perturbation power at fiber end, thus reducing
at each span the power-divergence problem intrinsic in the RP1 approximation
\cite{noi_RP}. Other redistribution mechanisms for which the above
theory applies verbatim are:

1) the thermally-induced guided-acoustic wave Brillouin scattering
(GAWBS)\cite{GAWBS}, for which $\delta P_{r}=\gamma_{GAWBS}\ell P$,
where $\gamma_{GAWBS}$ {[}km$^{-1}${]} is the GAWBS coefficient,
and $\ell$ {[}km{]} is the span length;

2) the inter-mode/core linear crosstalk in the MF, for which $\delta P_{r}=\gamma_{X}\ell P$,
where $\gamma_{X}$ {[}km$^{-1}${]} is the crosstalk coefficient
\cite{MMF_Xtalk}.

Thus including all three (uncorrelated) effects, we have in (\ref{eq:chir}),(\ref{eq:SNRr1}):
$\delta P_{r}=\alpha_{NL}P^{3}+(\gamma_{GAWBS}+\gamma_{X})\ell P$.

\section{\emph{\label{sec:SNR}}Signal to noise ratio}

According to the proposed per-span power-flow diagram in Fig. \ref{fig:block}(d),
the total span power gain seen by the transiting signal, i.e., the
overall span droop, is the product of addition and redistribution
droops: $\chi\triangleq\chi_{r}\chi_{a}.$ By the same reasoning as
in (\ref{eq:PsN}),(\ref{eq:PaN}), if the launch power is $P$, then
the desired multiplex signal power at the output of the $N$-th amplifier
is $P_{s}(N)=P\chi^{N}$ and therefore by the constant output power
constraint the accumulated \emph{addition+redistribution noise} at
the output of the N-span chain (over all modes and amplified WDM bandwidth
$B$) is $P_{a}(N)+P_{r}(N)=P(1-\chi^{N})$.

Hence the optical SNR (OSNR) at the output of the chain from amplifiers
1 to N, i.e., the ratio of total multiplex signal power to total noise
power at the output of the $N$-th amplifier, using (\ref{eq:chia}),(\ref{eq:chir})
is obtained as, 
\begin{equation}
OSNR=\frac{1}{\left[(1+SNR_{a1}^{-1})(1+SNR_{r1}^{-1})\right]^{N}-1}\label{eq:GDF}
\end{equation}
which is the generalized droop formula (GDF) derived for the first
time in \cite{ASN_droop_ofc19,ASN_suboptic} based on heuristic arguments.
The above derivation puts the GDF on a solid theoretical footing.
Please note a key assumption of the GDF model: the power additions
expressed by the power block diagram, Fig. \ref{fig:block}, tacitly
imply that the noise sources injected at each span are uncorrelated
from all others, and thus, in particular, assuming an incoherent accumulation
of NLI.

The GDF can be re-arranged into the key formula \cite{ASN_droop_ofc19,ASN_suboptic}
\begin{equation}
1+\frac{1}{OSNR}=\left[\left(1+\frac{1}{SNR_{a1}}\right)\left(1+\frac{1}{SNR_{r1}}\right)\right]^{N}\label{eq:PRID}
\end{equation}
which we call the \emph{product rule for inverse droop}. It hints
at the generalization \cite{ASN_droop_ofc19,ASN_suboptic}
\begin{equation}
1+\frac{1}{OSNR}=\prod_{k=1}^{N}\left(1+\frac{1}{SNR_{a1k}}\right)\left(1+\frac{1}{SNR_{r1k}}\right)\label{eq:PRID-1}
\end{equation}
for an inhomogeneous chain, where $SNR_{a1k}$ is the local ASE-reduced
OSNR at amplifier $k$, and similarly $SNR_{r1k}$ for redistribution
noise. We prove this generalization in Appendix A.

We conclude this section with a key observation. When the dominant
part of the power spectral density (PSD) of each of the above impairments
remains flat as the input signal PSD, then the per-tributary signal
to noise ratio $SNR$ for this flat-loss, flat-gain system will remain
equal to $OSNR$, since both signal and noises get filtered over the
same tributary bandwidth and mode. Hence from now on, we will drop
the ``O'' in the OSNR, and treat $P$ and $B$ as the launched input
power and bandwidth of each tributary. In section \ref{sec:The-failure-of}
we will generalize the per-tributary SNR expression to the case where
ASE occupies a larger bandwidth/number of modes than the signal multiplex.

\subsection*{SNR approximations}

We derive here upper and lower bounds to the GDF. Define 
\begin{equation}
SNR_{1}\triangleq(SNR_{a1}^{-1}+SNR_{r1}^{-1})^{-1}\label{eq:SNR1}
\end{equation}
as the SNR degraded by the total noise generated at a single span.
Let $x\triangleq SNR_{1}^{-1}$, which is normally a very small term.
Then, as proposed in \cite{ASN_droop_ofc19}, the GDF denominator
can be bounded as: $\chi^{-N}-1\geq(1+x)^{N}-1\geq Nx(1+\frac{1}{2}(N-1)x)$
by expanding to 2nd order in Taylor. Thus an upper-bound to the GDF
is 
\begin{equation}
SNR\leq\frac{SNR_{s}}{1+\frac{1}{2}(1-\frac{1}{N})(SNR_{s})^{-1}}\label{eq:snr_UB}
\end{equation}
where
\begin{equation}
SNR_{s}\triangleq\frac{1}{Nx}\equiv\frac{SNR_{1}}{N}\label{eq:SNR0}
\end{equation}
is the SNR we would calculate with the standard noise accumulation
formula for constant-gain amplifiers. We call it the \emph{standard
SNR}.

Now let $y=\frac{1}{2}(1-\frac{1}{N})Nx\geq0$. Since $(1+y)^{-1}\geq1-y$,
then we can lower-bound the upper-bound (\ref{eq:snr_UB}) and luckily
get a lower bound to the GDF-SNR as well:
\begin{equation}
SNR\geq\frac{1-\frac{1}{2}(1-\frac{1}{N})Nx}{Nx}=SNR_{s}-\frac{1}{2}(1-\frac{1}{N}).\label{eq:SNR_LB}
\end{equation}

\selectlanguage{english}%
\begin{figure}
\centering

\includegraphics[width=0.9\columnwidth]{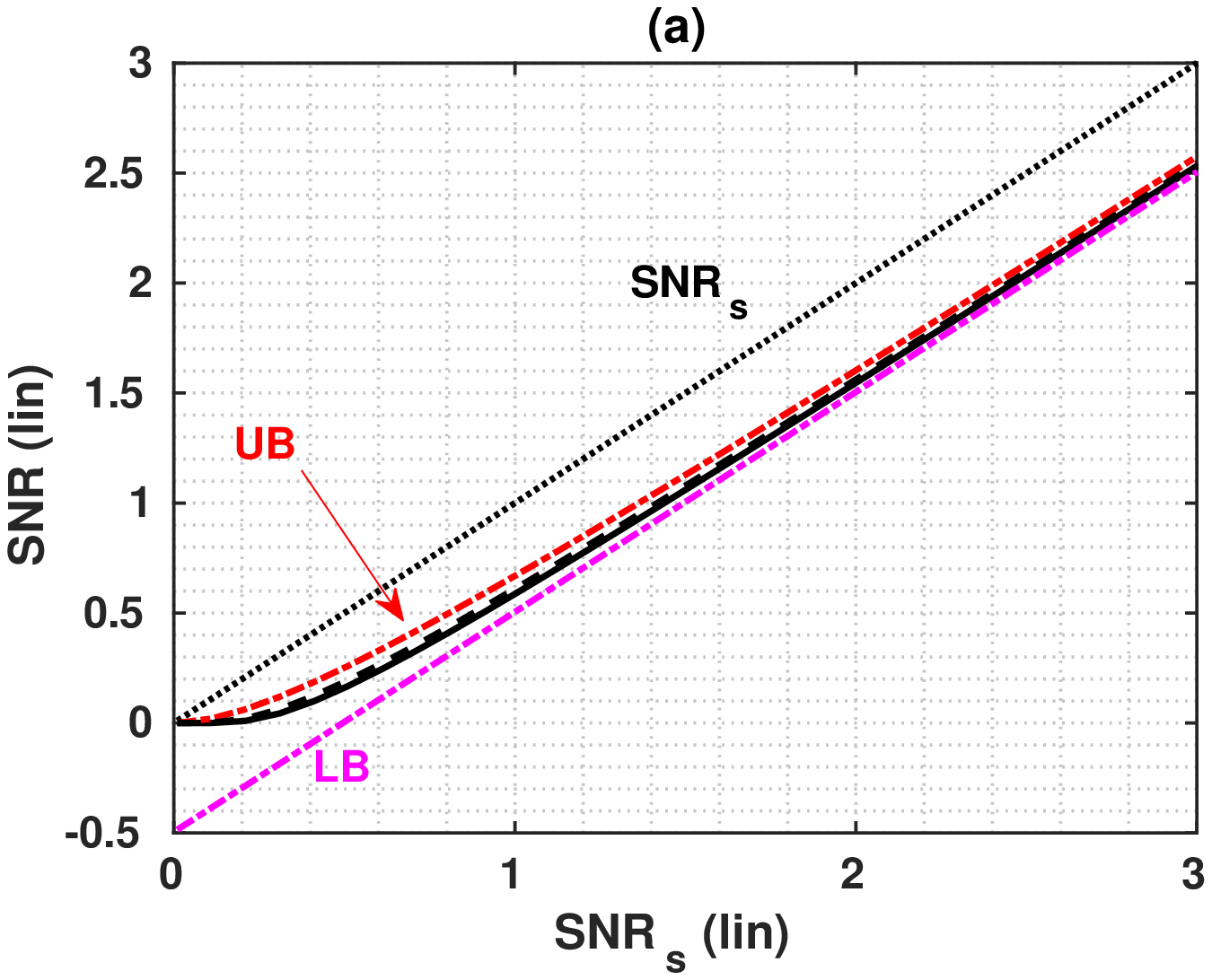}

\includegraphics[width=0.9\columnwidth]{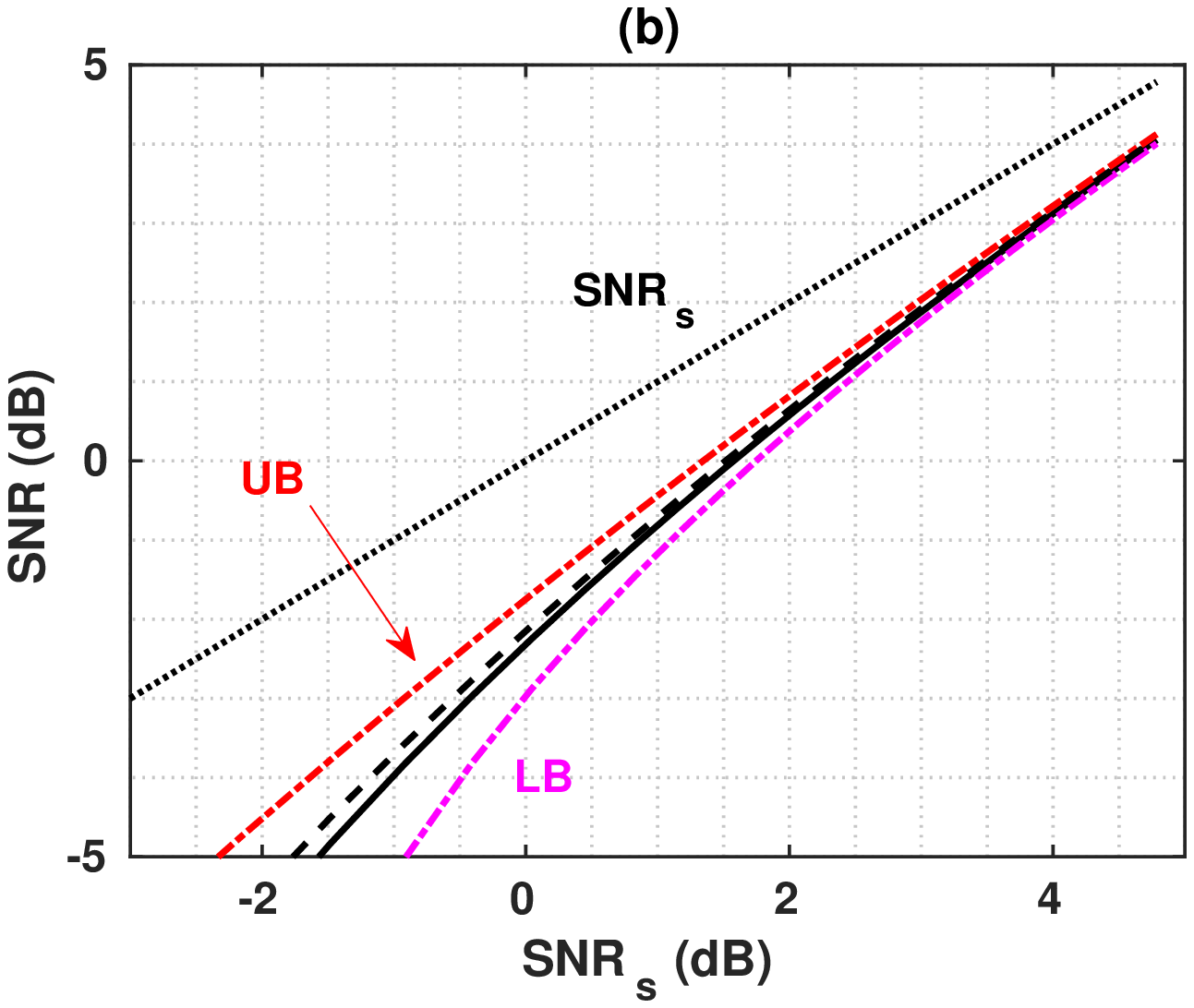}

\caption{\label{fig:SNR_UB_LB}(a) Solid: GDF (\ref{eq:GDF_SNR_appi}) versus
the standard SNR: $SNR_{s}=SNR_{1}/N$. Dash-dot: Upper-bound (UB)
eq. (\ref{eq:snr_UB}) and Lower-Bound (LB) eq. (\ref{eq:SNR_LB}).
Dotted: $SNR_{S}.$ Dashed: approximation (\ref{eq:SNR_app}). (b)
Corresponding figure when SNRs are in dB. Curves obtained for $N=100$,
but they remain essentially unchanged for any $N\protect\geq40$.}
\end{figure}

\selectlanguage{american}%
To understand the scope of the above approximations, Fig. \ref{fig:SNR_UB_LB}(a)
shows \foreignlanguage{english}{in black solid line} a plot of the
GDF-SNR\footnote{Since $SNR_{a1}^{-1}$ and $SNR_{r1}^{-1}$ are normally very small,
then the product $SNR_{a1}^{-1}SNR_{r1}^{-1}$ is a higher-order negligible
term. Hence the GDF (\ref{eq:GDF}) and expression (\ref{eq:GDF_SNR_appi})
are practically identical.}: \foreignlanguage{english}{
\begin{equation}
SNR\cong((1+SNR_{1}^{-1})^{N}-1)^{-1}\label{eq:GDF_SNR_appi}
\end{equation}
versus the standard $SNR_{s}\equiv SNR_{1}/N$. The figure also shows
its upper-bound (UB) eq. (\ref{eq:snr_UB}) and lower-bound (LB) eq.
(\ref{eq:SNR_LB}) (both dash-dotted), and the $SNR_{s}$ itself (dotted).
}Fig. \ref{fig:SNR_UB_LB}(b)\foreignlanguage{english}{ shows the
same as (a), but with SNRs expressed in dB. The curves in Fig. }\ref{fig:SNR_UB_LB}\foreignlanguage{english}{
were obtained for $N=100$, but they remain essentially unchanged
for any $N\geq40$.}

We observe in Fig. \ref{fig:SNR_UB_LB}(a)\foreignlanguage{english}{
and can prove analytically that: i) the LB (\ref{eq:SNR_LB}) crosses
zero at $SNR_{s}=\frac{N-1}{2N}$ and becomes negative (not physically
acceptable) below that; ii) the gap from $SNR_{s}$ to GDF (and to
all its approximations) converges for increasing $SNR_{s}$ to the
asymptotic value $\frac{1}{2}(1-\frac{1}{N})$, and the gap from $SNR_{s}$
to GDF exceeds 90\% of its asymptotic value at $SNR_{s}>1.66$ (2.2dB)
for any $N$. The constant gap in linear units translates into a variable
gap when SNRs are in dB, i.e., to a small dB-gap at large $SNR_{s}$
and a large dB-gap at lower $SNR_{s}$. The dB plot also does not
clearly show the $\sim1/2$ asymptotic linear gap. These observations
will be useful when interpreting the numerical SNR results in Section
\ref{sec:Numerical-checks}.}

\selectlanguage{english}%
Our best approximation to the GDF (which also turns out to be a tighter
upper-bound) is shown in dashed line in \foreignlanguage{american}{Fig.
\ref{fig:SNR_UB_LB}, and can be obtained by taking $10\log_{10}(.)$
of each side of eq. (\ref{eq:snr_UB}) and then linearizing the logarithm.
The resulting expression in dB is:
\begin{equation}
SNR(dB)\geq SNR_{s}(dB)-\frac{e^{dB}\cdot\frac{1}{2}(1-\frac{1}{N})}{SNR_{s}}\label{eq:SNR_app}
\end{equation}
where $SNR_{s}$ without (dB) indicates its linear value, $e^{dB}\triangleq10\log_{10}(e)\cong4.34$,
and $e$ is Neper's number.}

\selectlanguage{american}%
To make the physical meaning of the above approximations to the GDF
explicit, let's focus on the case of single-mode fibers ($M=1$),
with ASE and NLI only. Define 
\begin{equation}
\beta\triangleq h\nu FB\mathcal{L}^{-1}\label{eq:beta}
\end{equation}
as the ASE power (per mode) generated at the output of each amplifier
of gain $\mathcal{L}^{-1}$ over the per-tributary receiver bandwidth
$B$. Then from (\ref{eq:SNR1}),(\ref{eq:SNRa1}),(\ref{eq:SNRr1})
we get 
\begin{equation}
SNR_{s}\equiv\frac{1}{N(\frac{\beta}{P}+\alpha_{NL}P^{2})}\label{eq:GN}
\end{equation}
where in (\ref{eq:SNRa1}) we used $\delta P_{i}\mathcal{L}^{-1}=\beta$,
and the $\alpha_{NL}$ term is nominally the single-span coefficient
computed over the same per-tributary bandwidth $B$ as $\beta$ (more
on $\alpha_{NL}$ in Appendix B). We recognize the standard SNR (\ref{eq:GN})
to be the SNR of the well-known (incoherent) GN formula \cite{GN_poggio}.
The obtained upper and lower bounds are therefore approximations of
the GDF formula based solely on the value of the GN-SNR.

Although the above GN and GDF models assume uncorrelated NLI span
by span, in numerical computations we can approximately account for
the NLI span-by-span correlation by first calculating the NLI coefficient
of the entire link (using either the extended GN (EGN) model \foreignlanguage{english}{\cite{EGN1_torino,EGN2_torino,EGN_dar,EGN_serena}}
for the selected modulation format or by using SSFM simulations as
described in Appendix B) and then dividing by $N$, so that now the
$\alpha_{NL}$ to be used in the (coherent) GN\footnote{We should more correctly use the acronym EGN everywhere, since $\alpha_{NL}$
always accounts for the modulation format. We opted, however, to keep
GN since the term is more widespread.} and  GDF is a span-averaged coefficient and in general depends on
$N$.

\section{\label{sec:Numerical-checks}Numerical checks}

We present here three single-mode case studies with quasi Nyquist-WDM
signals where we verify the above formulas against SSFM simulations:

\textbf{case A)} is the 228x78km polarization-division multiplexed
(PDM) quadrature phase-shift keying (QPSK) WDM uncompensated link
analyzed in \cite{ASN_droop_ofc19}. The propagation fiber was \foreignlanguage{english}{an
EX2000$^{TM}$ (loss $0.169$ dB/km, }fiber nonlinear coefficient
$n_{2}=2.5\cdot10^{-20}$ m$^{2}$/W, effective area $110$ \textmu m\texttwosuperior ,
dispersion $20.7$ ps/nm/km). Optical amplifiers had a noise figure
$F$ of 8dB. The number of channels was 16, with channel spacing 37.5
GHz and symbol rate 34.17 Gbaud. SSFM simulations were carried out
with a simulated bandwidth 60 times the symbol rate, and ASE was removed
outside the WDM bandwidth. The number of transmitted symbols was $64800$.

\textbf{case B) }is the 190x78km PDM 16-quadrature amplitude modulation
(16QAM) link analyzed in \cite{ASN_droop_ofc19}. All data are the
same as in case A, except for the number of spans (now 190) and the
modulation format. The number of transmitted symbols was $2^{16}$.

\textbf{case C)} is the 40x120km PDM-QPSK link analyzed in \cite[Fig. 3]{amir_droop}.
The propagation fiber was \foreignlanguage{english}{a non-zero dispersion
shifted fiber (NZDSF) (loss $0.22$ dB/km, }fiber NL coefficient $n_{2}=2.6\cdot10^{-20}$
m$^{2}$/W, effective area $70.26$ \textmu m\texttwosuperior , dispersion
$3.8$ ps/nm/km). Noise figure $F$ was 5dB. The number of channels
was 15, with channel spacing 50 GHz and symbol rate 49 Gbaud. Again
the simulated bandwidth was 60 times the symbol rate, and ASE was
removed outside the WDM bandwidth. The number of transmitted symbols
was $2^{13}$.

\selectlanguage{english}%
\begin{figure}
\centering

\includegraphics[width=1\columnwidth]{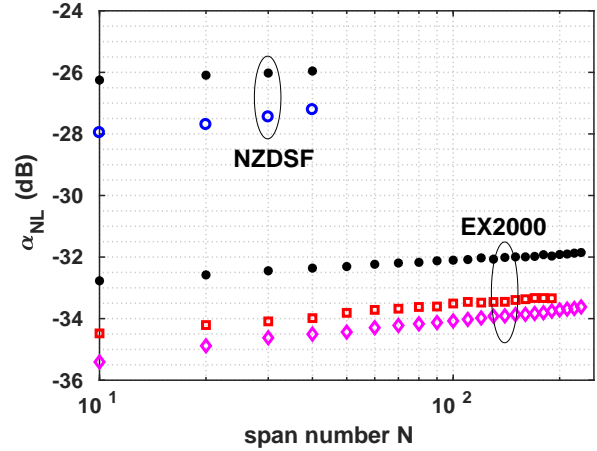}

\caption{\label{fig:sim_aNL_vs_N} \foreignlanguage{american}{Values in dB
of span-averaged NLI coefficient $\alpha_{NL}$ (mW$^{-2}$) versus
spans $N$, used in theoretical formulas for the 3 case studies (diamonds
case A, squares case B, circles case C). Also, filled circles show
values for Gaussian modulation.}}
\end{figure}

\selectlanguage{american}%
We accounted just for ASE and NLI\foreignlanguage{english}{, and the
GDF formula explicitly is
\begin{equation}
SNR=\frac{1}{\left[\left(1+\frac{\beta}{P}\right)\left(1+\alpha_{NL}P^{2}\right)\right]^{N}-1}.\label{eq:GDF_11}
\end{equation}
}

For all 3 cases, we will present results at the stated number of spans,
as well as some results at lower span numbers, all multiples of 10.
Fig. \ref{fig:sim_aNL_vs_N} shows the values of the span-averaged
$\alpha_{NL}$ we have used in the theoretical formulas in the 3 cases
(diamonds for case A, squares for case B, circles for case C), along
with the values for Gaussian modulation (filled circles). We note
in passing that an $\alpha_{NL}$ that grows with $N$ is an indication
of self-nonlinearities becoming more important than cross-nonlinearities
as $N$ increases, which is typical of small WDM systems \cite{Vacondio}.

\selectlanguage{english}%
\begin{figure*}
\centering\includegraphics[width=0.7\columnwidth]{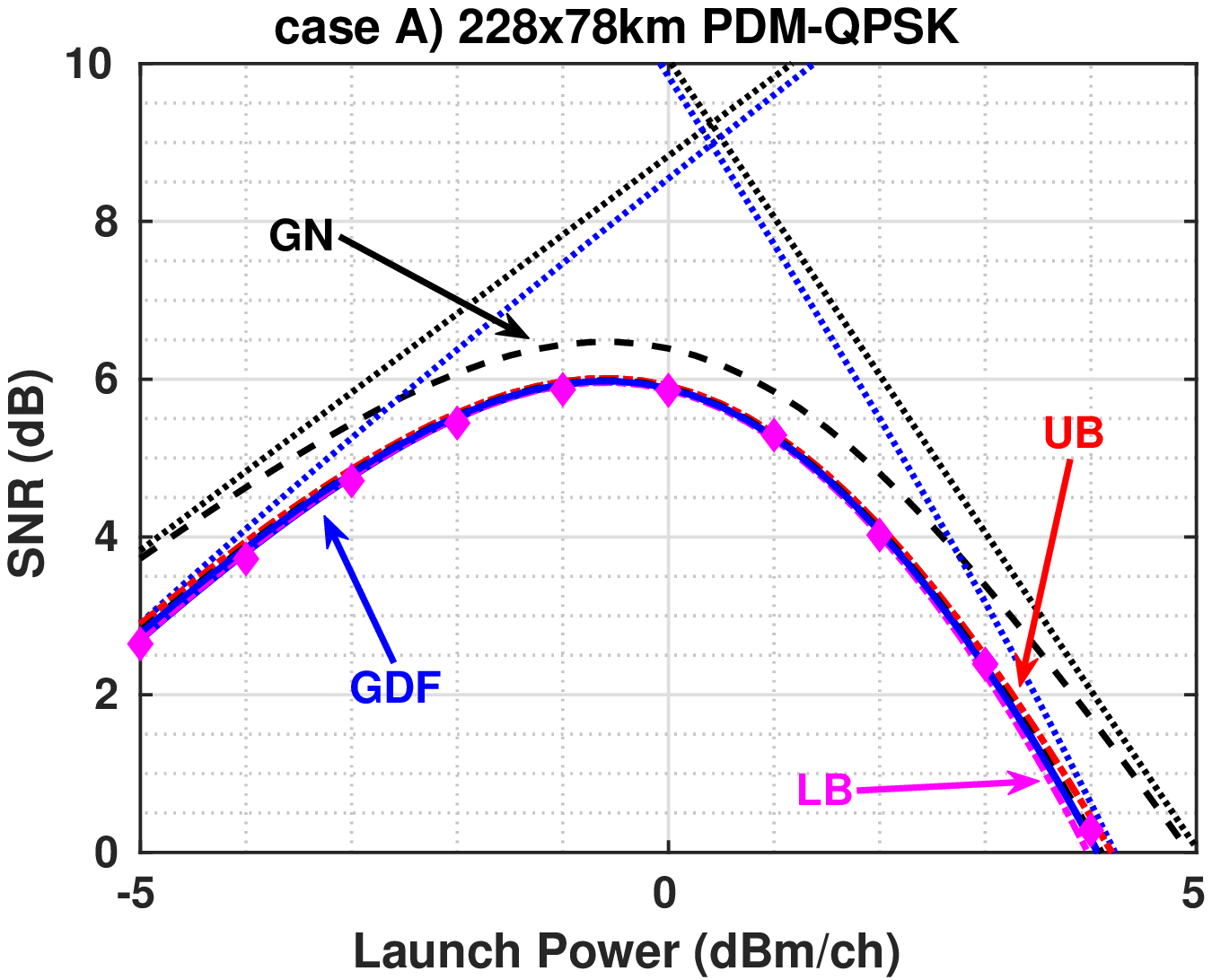}\includegraphics[width=0.7\columnwidth]{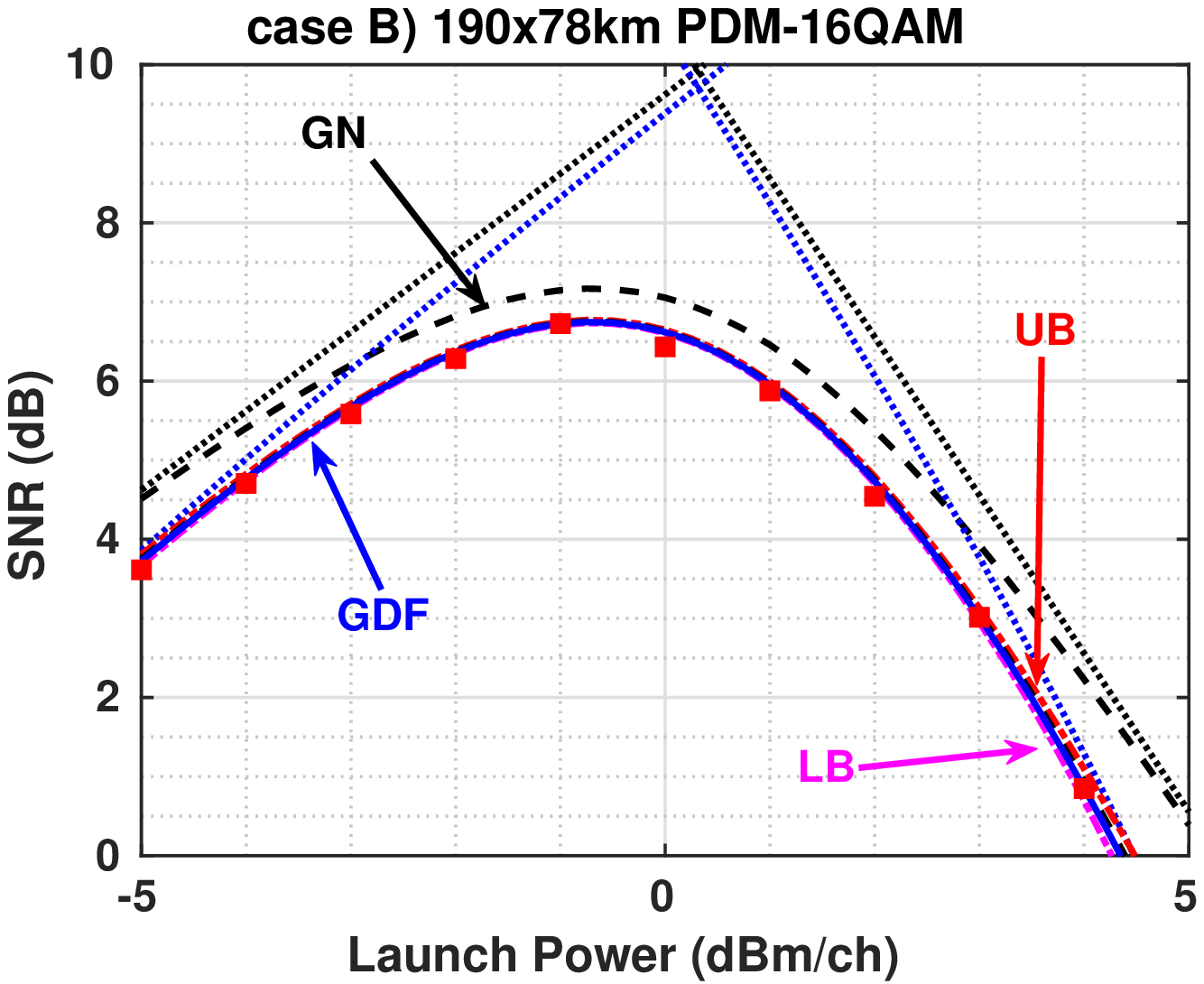}\includegraphics[width=0.7\columnwidth]{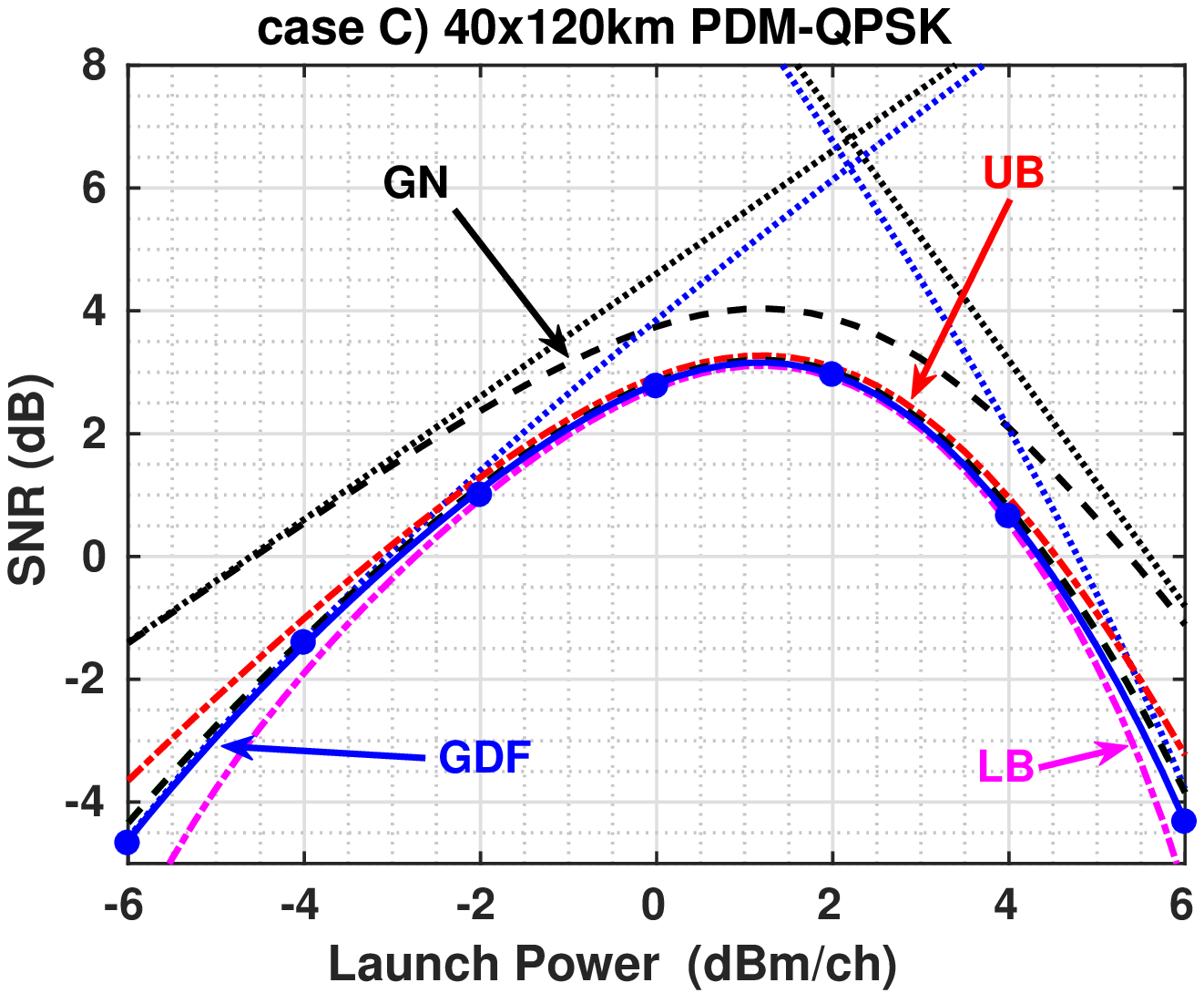}

\caption{\label{fig:SNR_vs_P} SNR(dB) versus power per channel $P$(dBm) for:
\textbf{case A)} 228x78km EX2000 uncompensated link, 34.17 Gbd PDM-QPSK,
16-channel @ 37.5GHz spacing \cite{ASN_droop_ofc19}; \textbf{case
B)} 190x78km EX2000 uncompensated link, 34.17 Gbd PDM-16QAM, 16-channel
@ 37.5GHz \cite{ASN_droop_ofc19}; \textbf{case C)} 40x120km NZDSF
uncompensated link, 49-Gbd PDM-QPSK, 15-channel @ 50GHz \cite{amir_droop}.
In each plot: Symbols: SSFM simulations. Blue solid: GDF eq. (\ref{eq:GDF_11}).
Black dashed: GN formula eq. (\ref{eq:GN}). Linear and nonlinear
asymptotes also shown in dotted lines. GDF Approximations: Red dash-dotted:
upper-bound (UB) (\ref{eq:snr_UB}); Magenta dash-dotted: lower-bound
(LB) (\ref{eq:SNR_LB});\foreignlanguage{american}{ }black dashed:
approximation (\ref{eq:SNR_app})\foreignlanguage{american}{.}}
\end{figure*}

We begin by presenting in Fig. \ref{fig:SNR_vs_P} the received SNR
versus transmitted power per channel $P$ for all 3 cases at their
maximum distance. In all 3 sub-figures we report: the GN formula (\ref{eq:GN})
(dashed black) and the GDF (\ref{eq:GDF_11}) (solid blue), along
with their linear and nonlinear asymptotes (dotted); the SSFM simulations
at constant output power (symbols: diamonds for case A, squares for
case B, circles for case C); the upper and lower bounds UB (\ref{eq:snr_UB}),
LB (\ref{eq:SNR_LB}) (both dash-dotted), and the approximation (\ref{eq:SNR_app})
(dashed). \foreignlanguage{american}{Values of $\alpha_{NL}$ estimated
from low-power simulations and used in theoretical formulas were:
$\alpha_{NL}=[4.34,\,\,4.63,\,\,19.01]\times10^{-4}$ (mW$^{-2}$)
for cases A,B,C, respectively. These can be read from Fig. \ref{fig:sim_aNL_vs_N}.}

We first note in all cases the very good fit of the GDF with SSFM
simulations, not only in cases A and B (simulations here are more
accurate than those in \cite{ASN_droop_ofc19}, which, however, already
showed a good match), but also in case C, where the end-to-end RP1
model developed in \cite{amir_droop} was unable to match the simulated
SNR at very large powers where NLI-induced droop is significant (Cfr.
\cite{amir_droop} Fig. 5, label ``Unm.''); the local-RP1 power-renormalized
concatenation implicit in the GDF is instead able to  well reproduce
the SNR even in deep ``signal depletion'' by NLI. Note also that,
although the GDF assumes uncorrelated noises at each span, our use
of the span-averaged NLI coefficient allows us to have good fit also
in links where span-by span correlations are significant, as in our
3 cases where Fig. \ref{fig:sim_aNL_vs_N} shows a marked variation
of $\alpha_{NL}$ with span number $N$.

Next note that the GDF is always below the GN and its asymptotes have
a different slope than those of the GN. As already noted in \foreignlanguage{american}{Fig.
\ref{fig:SNR_UB_LB},} this is an artifact of the dB representation
of the SNR, since the SNR gap between the two curves is about $1/2$
(in linear units) for GN-SNR above $\sim$1.66 (2.2 dB).

Finally, we note that UB and LB and approximation (\ref{eq:SNR_app})
are basically coinciding with the GDF in cases A and B on the shown
scale, and they become visible in the tails of the SNR ``bell-curve''
in case C. The gap to UB, LB and (\ref{eq:SNR_app}) is quite small,
as already seen in Fig. \ref{fig:SNR_UB_LB}. In particular, approximation
(\ref{eq:SNR_app}) is the best among all, and is basically coinciding
with the GDF over most of the shown ranges.

\begin{figure}
\centering

\includegraphics[width=1\columnwidth]{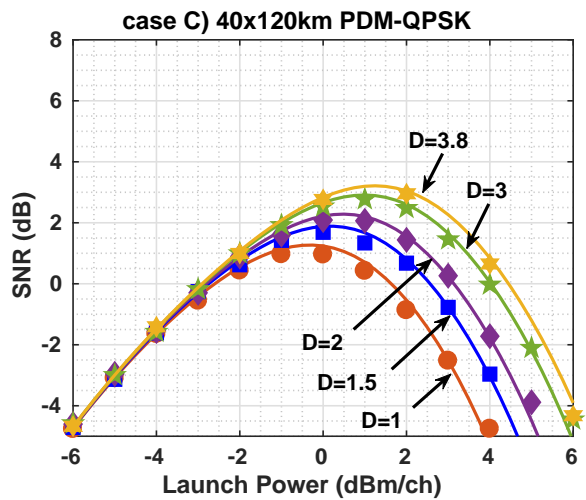}

\caption{\label{fig:amir_D} \foreignlanguage{american}{Case C) SNR versus
launch power when lowering fiber dispersion $D$ from 3.8 down to
1 ps/nm/km. Symbols: simulations. Solid: GDF.}}
\end{figure}

To test the resilience of the GDF at low dispersion, Fig. \ref{fig:amir_D}
shows, for the same parameters as case C except dispersion, the GDF-SNR
versus launch power (solid), along with SSFM simulations (symbols),
when dispersion is decreased \foreignlanguage{american}{from 3.8 down
to 1 ps/nm/km}. We see that the GDF-SNR well reproduces the simulated
SNR over the whole power range down to dispersions of 1 ps/nm/km.

\subsection{\label{subsec:Comparisons-with-the}Comparisons with the GN formula}

Since the GN is the reference formula for nonlinear propagation with
coherent detection, it is important to quantify its gap in performance
to the GDF.
\selectlanguage{american}%

\subsubsection*{SNR gap}

\selectlanguage{english}%
\begin{figure}
\centering

\includegraphics[width=1\columnwidth]{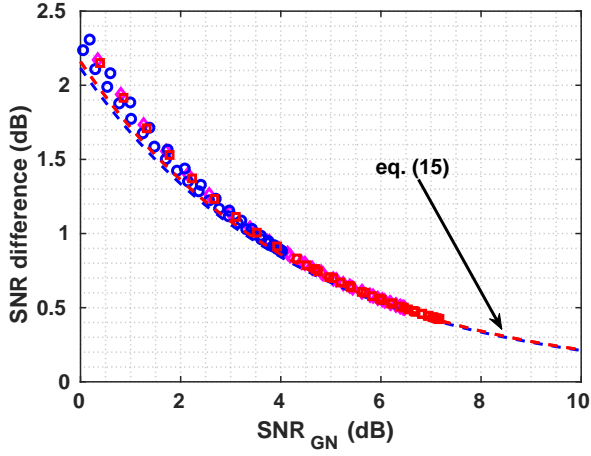}

\caption{\label{fig:DSNR_vs_GN} Gap $SNR_{GN}(dB)-SNR_{GDF}(dB)$ versus $SNR_{GN}(dB)$.
Symbols (diamonds for case A, squares case B, circles for case C):
exact gap, as visible in Fig. \ref{fig:SNR_vs_P}. Dashed: approximation
(\ref{eq:SNR_app}).}
\end{figure}

We here discuss the gap from $SNR_{GN}$ eq. (\ref{eq:GN}) to $SNR_{GDF}$
eq. (\ref{eq:GDF_11}). The bounds we have found all hint at a 1-1
relation between the two SNRs. This is not exactly so, but almost.
Fig. \ref{fig:DSNR_vs_GN} plots the gap $SNR_{GN}(dB)-SNR_{GDF}(dB)$
versus $SNR_{GN}(dB)$. Symbols for the three cases (diamonds for
case A, squares for case B and circles for case C) indicate the exact
gap between the theoretical SNRs (we measure the gap from Fig. \ref{fig:SNR_vs_P}
in each case scanning from low to high power, and report the values
in Fig. \ref{fig:DSNR_vs_GN}), while the dashed lines indicate the
gap as expressed by the best approximation (\ref{eq:SNR_app}). Especially
for case C (circles) it is evident that the $SNR_{GDF}$ is not a
1-1 function of $SNR_{GN}$, but to a good extent we may well approximate
the gap for all systems by eq. (\ref{eq:SNR_app}). The gap does not
exceed 2.5dB for $SNR_{GN}$ down to 0 dB.

\subsubsection*{Optimal power at max SNR}

The optimal power $P_{o}$ at maximum SNR is obtained in the GN model
by setting the derivative of $SNR_{GN}$ w.r.t. $P$ to zero, yielding
the condition $\beta=2\alpha_{NL}P_{o}^{3}$ (i.e., ASE is twice the
NLI at $P_{o}$) and the explicit optimal GN power $P_{oGN}=(\beta/2/\alpha_{NL})^{1/3}$.

Similarly, the GDF-SNR is maximum at the power $P_{o}$ that makes
the total droop $\chi(P_{o})$ closest to 1, leading to the condition
$\beta=\frac{2}{\chi(P_{o})}\alpha_{NL}P_{o}^{3}$ , i.e., ASE is
\emph{slightly more than twice} the NLI at $P_{o}$. This leads to
$P_{o}=P_{oGN}\chi^{1/3}\lesssim P_{oGN}$, since the droop per span
$\chi=\chi_{a}\chi_{r}$ is always practically very close to 1. Thus
the optimal $P_{o}$ for the GDF is in practice the same as in the
GN case,
\selectlanguage{american}%

\subsubsection*{Spectral efficiency per mode}

\selectlanguage{english}%
A lower-bound on the capacity per mode of the nonlinear optical channel
for dual-polarization transmissions is obtained from the equivalent
additive white Gaussian noise (AWGN) Shannon channel capacity, i.e.,
by considering the NLI as an additive white Gaussian process independent
of the signal. Hence a lower-bound on spectral efficiency per mode
is \cite{GN_poggio,TEsubcom_jlt18}: $SE=2\log_{2}(1+SNR)$ {[}b/s/Hz{]}.
Its top value $SE_{o}$ is achieved at $P_{o}$ using its corresponding
top SNR.
\begin{figure}
\centering\includegraphics[width=1\columnwidth]{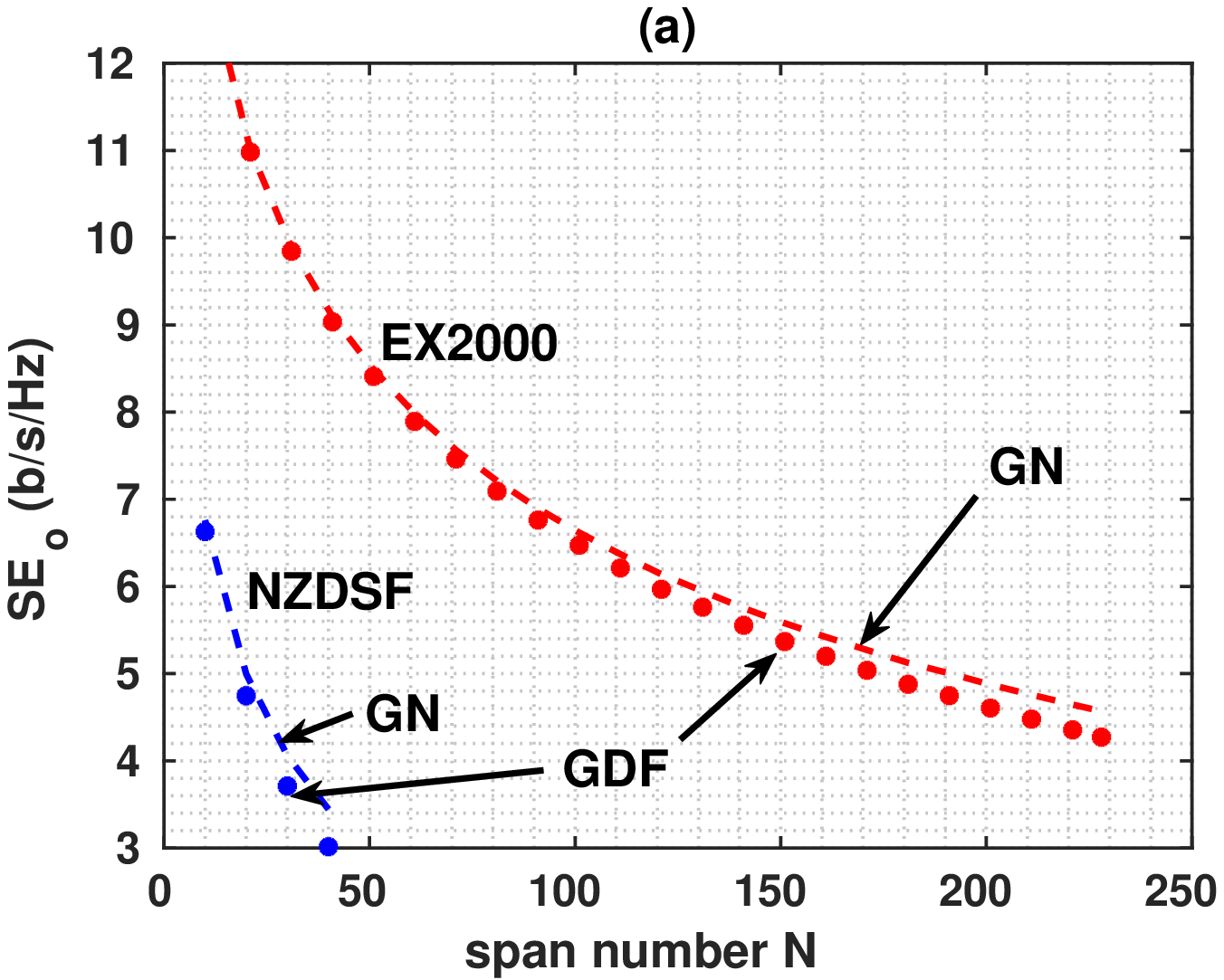}

\includegraphics[width=1\columnwidth]{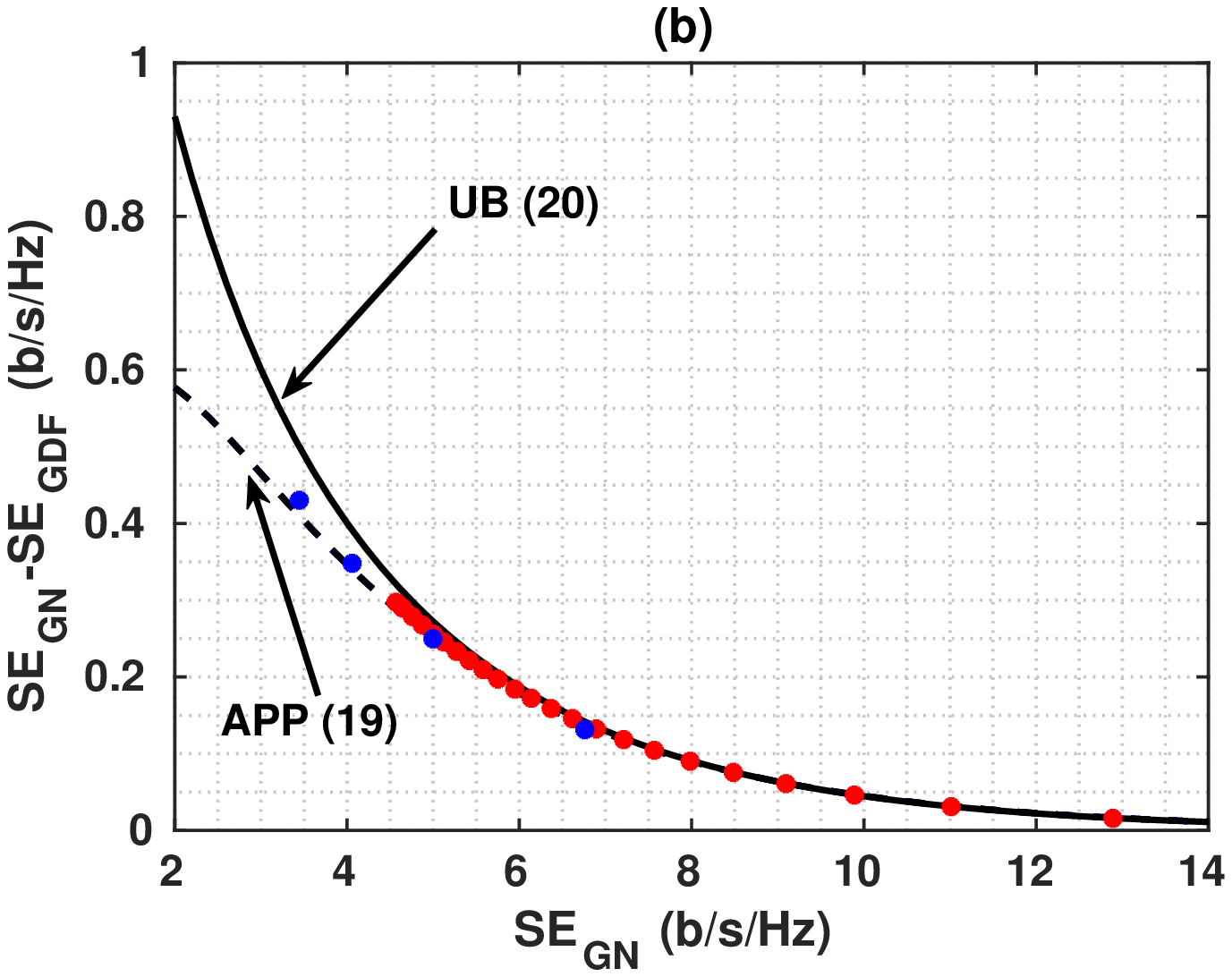}

\caption{\label{fig:SEvsN} (a) (lower-bound to) top spectral efficiency $SE_{0}$
(b/s/Hz) versus span number $N$, obtained for the EX2000 (case A/B)
and the NZDSF (case C) links with Gaussian modulation. Dashed: GN
model. Filled circles: GDF. (b) Filled circles: gap from GN to GDF
SE versus GN-SE. Solid: UB (\ref{eq:DSE_UB}). Dashed: approximation
(\ref{eq:DSE}).}
\end{figure}

\selectlanguage{american}%
For fixed distance, symbol rate, and noise figure, the GDF- and GN-SNR
just depend on the NLI per-span parameter $\alpha_{NL}$. The $\alpha_{NL}$
values for the AWGN capacity-achieving Gaussian modulation are reported
with filled circles in Fig. \ref{fig:sim_aNL_vs_N} for both the EX2000
and the NZDSF links.

\selectlanguage{english}%
Fig. \ref{fig:SEvsN}(a) reports (a lower-bound to) the top $SE_{0}$
versus span number $N$ for both GN (dashed) and GDF (filled circles)
obtained for the EX2000 and the NZDSF links with Gaussian modulation.
The figure shows that a noticeable departure of the correct GDF-SE
from the GN-SE occurs only at GN-SE values below 5 b/s/Hz.

We prove in Appendix C that the SE gap from GN to GDF $\Delta SE\triangleq SE_{GN}-SE_{GDF}$
is well approximated at large $N$ and at all powers (not only at
top) by 
\begin{equation}
\Delta SE\cong\frac{2}{\ln(2)}\frac{SNR_{GN}}{1+2SNR_{GN}+2SNR_{GN}^{2}}\label{eq:DSE}
\end{equation}
and upper bounded by
\begin{equation}
\Delta SE\leq\frac{1}{\ln(2)[SNR_{GN}+\frac{1}{2}]}\label{eq:DSE_UB}
\end{equation}
which are plotted in Fig. \ref{fig:SEvsN}(b) versus $SE_{GN}$ in
dashed and solid line, respectively, together with the exact gap (filled
circles). Curve (\ref{eq:DSE}) can be taken as a good approximation
to all shown cases. From the figure, it is seen that the GN model
over-estimates SE by less than 0.35 {[}b/s/Hz{]} at SE predicted by
the GN model above 4 {[}b/s/Hz{]}, and over-estimates SE by between
0.6 and 0.9 {[}b/s/Hz{]} at $SE_{GN}=2$ {[}b/s/Hz{]}.

\section{\label{sec:The-failure-of}Limits of the GDF}

\begin{figure}
\centering

\includegraphics[width=1\columnwidth]{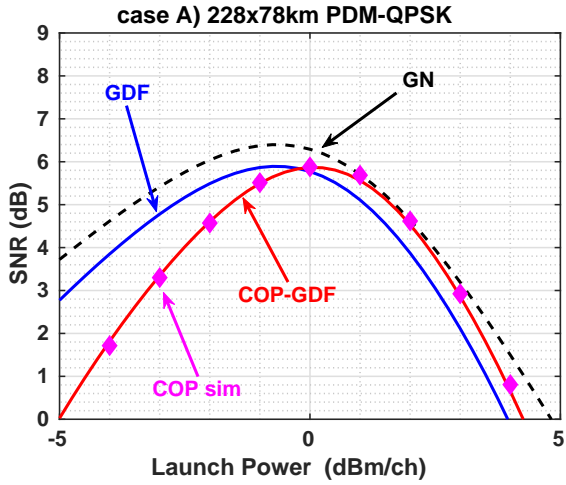}

\caption{\label{fig:caseA_EGDF} \foreignlanguage{american}{Case A) SNR versus
launch power when amplifier bandwidth is $B_{a}=60B_{rx}$, $B_{rx}=34.17$
GHz, and ASE unfiltered. 16 PDM-QPSK channels, spacing $\Delta f=37.5$
GHz. Symbols: simulations with COP saturation power $16P_{t}$. We
show: the GN (\ref{eq:GN}) and the raw GDF (}\ref{eq:GDF_11}\foreignlanguage{american}{)
(where $B\to B_{rx}$ and $P\to P_{t}$); and the COP-GDF, eqs. (\ref{eq:eGDF}),
(\ref{eq:chi_COP-GDF}).}}
\end{figure}

Unfortunately, when the amplified modes $M_{a}$ and amplified bandwidth
$B_{a}$ exceed the signal modes $M$ and signal occupied bandwidth
$B=N_{c}B_{rx}$ (possible gaps between WDM channels are not counted),
the GDF ceases to be accurate, and the \foreignlanguage{american}{\emph{amplifier
fill-in efficiency}: 
\begin{equation}
\eta_{A}\triangleq\frac{MN_{c}}{M_{a}N_{a}}\label{eq:etaA}
\end{equation}
with $N_{a}\triangleq B_{a}/B_{rx}$, plays a major role in setting
performance.}

\selectlanguage{american}%
As a numerical example, we consider the 228x78km 16-channel single-mode
($M=M_{a}=1$) PDM-QPSK case study A, but now ASE is present over
the whole amplified (and simulated) bandwidth $B_{a}=60B_{rx}$, with
$B_{rx}=34.17$ GHz. In this system we have $\eta_{A}=0.266$. This
small number should be checked against the value $\eta_{A}=0.91$
for case A in Fig. \ref{fig:SNR_vs_P}, where ASE is filtered over
the WDM bandwidth and the basic GDF very well matches simulations.

Fig. \ref{fig:caseA_EGDF} shows the per-tributary SNR versus launch
power per tributary $P_{t}$ (saturation power is in general $P=MN_{c}P_{t}$).
Symbols are SSFM simulations. We also see in solid line the GN (\ref{eq:GN})
and the raw GDF (\foreignlanguage{english}{\ref{eq:GDF_11}}) (where
in the referenced equations we set $B\to B_{rx}$ and $P\to P_{t}$).
We note that with a low $\eta_{A}$ the GDF ceases to well match the
simulations. This is mostly due to the fact that, because of the relevant
out-of-band ASE, the actual ASE-droop is larger and the actual NLI-droop
is smaller than what the GDF predicts. \foreignlanguage{english}{The
next sub-section explains the tricks necessary to modify the basic
GDF to cope with such a scenario.}

\subsection{The COP-GDF}

Let $P_{s}(N),P_{a}(N),P_{r}(N)$ be the total cumulated signal, ASE
and NLI redistribution power from the link input up to the output
of span $N$. Assuming equal per-tributary powers, \foreignlanguage{english}{at
the per-tributary receiver the SNR is: 
\begin{align}
SNR & =(\frac{P_{s}(N)}{MN_{c}})/(\frac{P_{a}(N)}{M_{a}N_{a}}+\frac{P_{r}(N)}{MN_{c}})\nonumber \\
 & =\frac{P_{s}(N)}{P_{a}(N)\eta_{A}+P_{r}(N)}\label{eq:tributarySNR}
\end{align}
where we assumed that NLI is the same at all tributaries and exists
only over the same modes/spectral range as the signal multiplex. It
is evident from (\ref{eq:tributarySNR}) that SNR evaluation, differently
from the basic GDF (\ref{eq:GDF}), now requires a separate evaluation
of both $P_{a}(N)$ and $P_{r}(N)$. These can be calculated explicitly
as shown in (\ref{eq:SOLrecursions}) in Appendix A, yielding
\begin{equation}
SNR=\frac{\prod_{m=1}^{N}\chi_{m}}{\sum_{k=1}^{N}[(\chi_{a}^{-1}-1)\chi_{rk}^{-1}\eta_{A}+\chi_{rk}^{-1}-1]\prod_{m=k}^{N}\chi_{m}}\label{eq:eGDF}
\end{equation}
which requires an explicit evaluation of the ASE droop $\chi_{a}$
(\ref{eq:chia-1}) and of the redistribution droop $\chi_{rk}$ (\ref{eq:chir-1}).}

Specifically, regarding $\chi_{a}=(1+SNR_{a1}^{-1})^{-1}$, this is
span-independent, since $\delta P_{i}=M_{a}N_{a}h\nu FB_{rx}$, so
that 
\begin{align}
SNR_{a1} & \triangleq\frac{P}{\delta P_{i}\mathcal{L}^{-1}}=\frac{MN_{c}P_{t}}{M_{a}N_{a}h\nu FB_{rx}\mathcal{L}^{-1}}\nonumber \\
 & =\eta_{A}P_{t}/\beta\label{eq:SNRa1-2}
\end{align}
where we used definition of $\beta$ (eq.(\ref{eq:beta}) where $B\to B_{rx}$),
and the definition of $\eta_{A}$ (\ref{eq:etaA}).

Regarding the NLI droop $\chi_{rk}=(1+SNR_{r1k}^{-1})^{-1}$ in eq.
(\ref{eq:chir-1}), this is now span-dependent because, due to the
COP constraint, the out-of-band/mode ASE (O-ASE) reduces the \emph{effective}
tributary power $P_{e}$ that generates NLI, more and more as the
spans increase.

To find the correct per-tributary effective power $P_{e}(k)$ generating
NLI at span $k$ we reason as follows. At each span $k=1,..,N$, the
total power that effectively contributes to the NLI generation is
not $P=MN_{c}P_{t}$, but $P$ minus the O-ASE power entering span
$k$, $P_{ASE,O}(k)$, which we now calculate.

The locally generated output O-ASE at each amplifier is $\beta'=\beta N_{a}(M_{a}-M)+\beta(N_{a}-N_{c})M$,
i.e., the sum of the whole ASE over non-signal modes and the out-of-band
ASE on signal modes. With our definitions, this simplifies to $\beta'=\beta(\frac{1}{\eta_{A}}-1)MN_{c}$.
Hence the cumulated (and drooped) O-ASE up to span $k$ is
\[
P_{ASE,O}(k)\triangleq\beta'(\chi_{2}\cdot\cdot\chi_{k-1}+\chi_{3}\cdot\cdot\chi_{k-1}+..+\chi_{k-1}+1)
\]
where the final $1$ is due to the O-ASE generated at amplifier $k-1$
which is not drooped. By approximating each droop as just the ASE
droop: $\chi_{j}\cong\chi_{a}$ we thus finally get the effective
power and the single-span nonlinear SNR as
\begin{equation}
\left\{ \begin{array}{l}
P_{e}(k)=\frac{P-P_{ASE,O}(k)}{MN_{c}}\cong P_{t}-\beta(\frac{1}{\eta_{A}}-1)\frac{1-\chi_{a}^{k-1}}{1-\chi_{a}}\\
SNR_{r1k}=\frac{P}{\delta P_{rk}}\equiv\frac{P_{t}}{\alpha_{NL}P_{e}(k)^{3}}
\end{array}\right..\label{eq:P_e}
\end{equation}

In summary, the resulting improved SNR formula, which we call the
\emph{COP-GDF}, is calculated by eq.\foreignlanguage{english}{ (\ref{eq:eGDF}),
where using (\ref{eq:SNRa1-2}) and (\ref{eq:P_e}) we have
\begin{equation}
\left\{ \begin{array}{l}
\chi_{a}^{-1}=1+\beta/(\eta_{A}P_{t})\\
\chi_{rk}^{-1}\cong1+\alpha_{NL}P_{t}^{2}\left(1-\frac{\beta}{P_{t}}(\frac{1}{\eta_{A}}-1)\frac{1-\chi_{a}^{k-1}}{1-\chi_{a}}\right)^{3}
\end{array}\right..\label{eq:chi_COP-GDF}
\end{equation}
}

For case study A, Fig. \ref{fig:caseA_EGDF} also reports the COP-GDF
and shows that it well matches simulations. Similar results are obtained
for case B and are not reported.

\selectlanguage{english}%
\begin{figure}
\centering

\includegraphics[width=1\columnwidth]{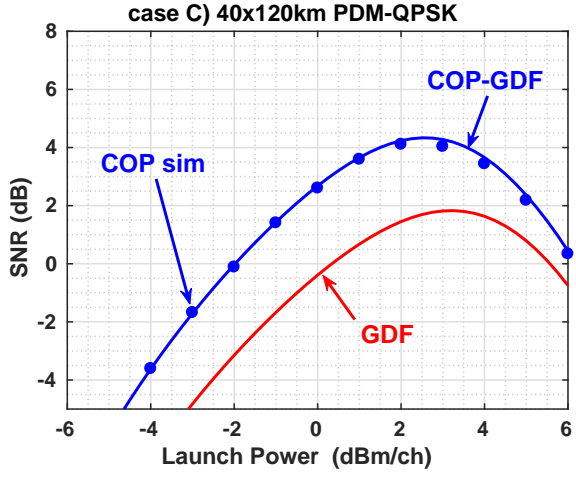}

\caption{\label{fig:caseC_100G} \foreignlanguage{american}{Case C) SNR versus
launch power $P_{t}$, Cfr. Fig. \ref{fig:SNR_vs_P} (15-channel 49Gb
PDM-QPSK 40x120km NZDSF link with $\eta_{A}=0.98$). ASE filtered
on WDM bandwidth, but now channel spacing has doubled to $\Delta f=100$
GHz, yielding an amplifier fill-in efficiency $\eta_{A}=0.49$. Symbols:
simulations with COP saturation power $15P_{t}$. We show: the basic
GDF (}\ref{eq:GDF_11}\foreignlanguage{american}{) where $B\to B_{rx}=49GHz$
and $P\to P_{t}$, and the COP-GDF, eq. (\ref{eq:eGDF}),(\ref{eq:chi_COP-GDF}).}}
\end{figure}

\selectlanguage{american}%
We show in Fig. \ref{fig:caseC_100G} for case study C what happens
when channel spacing is doubled to 100 GHz with respect to Fig. \ref{fig:SNR_vs_P},
and thus the amplifier fill-in efficiency is halved to $\eta_{A}=0.49$.
We see that the basic GDF formula badly fails, but the COP-GDF well
matches the SSFM simulations.

\section{\label{sec:The-constant-gain-case}The constant-gain case}

For CG amplifiers, the extension of the GN model to include the nonlinear
signal-noise interaction, and its induced signal power depletion,
was tackled in \cite{serena_S-NLI_interaction} with a rigorous end-to-end
RP1 model (which inspired the COP-amplifier end-to-end RP1 model in
\cite{amir_droop}), and heuristically in \cite{poggio_depletion-2}.
While the RP1 model in \cite{serena_S-NLI_interaction} has the same
intrinsic inability as the model in \cite{amir_droop} to cope with
the NLI-induced droop at large power, the heuristic models in \cite{poggio_depletion-2}
do go beyond the RP1 limits. We next review such models, and introduce
our new contribution, namely the CG-GDF, that uses the Turin's group
physical intuitions \cite{poggio_depletion-2} to extend the COP-GDF
ideas and yield a very accurate SNR estimation formula even for the
CG case.

The CG-SNR formulas in \cite{poggio_depletion-2} are the following:

1) \emph{TU}: it calculates the formula $SNR=\frac{P_{t}}{N\beta+P_{NLI}}$
with 
\begin{equation}
P_{NLI}=\alpha_{NL}\sum_{n=0}^{N-1}(P_{t}+n\beta)^{3}\label{eq:PNLI_carena}
\end{equation}
that approximately accounts for ASE-signal nonlinear interaction by
suitably modifying the estimated NLI variance \cite[eq. (5)]{poggio_depletion-2}\footnote{Summation in \cite{poggio_depletion-2} runs 1 to $N$, but in our
simulations the launched power is without ASE, hence we sum 0 to $N-1$.}. Note that this formula assumes that the ASE useful for nonlinear
calculations is the one over the signal bandwidth, since O-ASE is
ineffective for CG amplifiers.

2) \emph{TL}: it calculates the formula
\begin{equation}
SNR=\frac{P_{t}-P_{NLI}}{N\beta+P_{NLI}}\label{eq:carena_depletion}
\end{equation}
with $P_{NLI}$ as in (\ref{eq:PNLI_carena}).

\selectlanguage{english}%
\begin{figure}
\centering

\includegraphics[width=1\columnwidth]{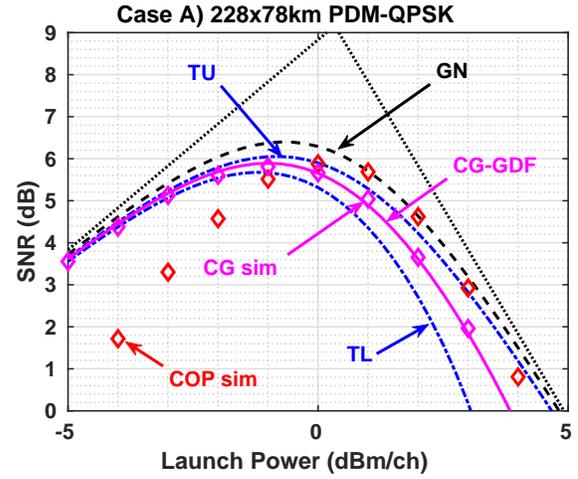}

\caption{\label{fig:CG_finale} \foreignlanguage{american}{Case A) SNR versus
launch power $P_{t}$ when amplifier bandwidth is $B_{a}=60\Delta f$,
$\Delta f=37.5$ GHz, and ASE unfiltered, as in Fig. \ref{fig:caseA_EGDF}.
Symbols: simulations for both CG and COP. We show: the GN formula
(dash); TU and TL formulas, eq. (\ref{eq:PNLI_carena}),(\ref{eq:carena_depletion})
(dash-dot); and the CG-GDF, eq. (\ref{eq:eGDF}),(\ref{eq:chi_CG-GDF}).}}
\end{figure}

\selectlanguage{american}%
For case study A), Fig. \ref{fig:CG_finale} reports the tributary
SNR versus launch power $P_{t}$ for both CG and COP amplifiers, when
amplifier bandwidth is $B_{a}=60B_{rx}$ and ASE is unfiltered, same
as in Fig. \ref{fig:caseA_EGDF}. Symbols are the simulations for
both CG and COP amplifiers, the dashed curve is the GN formula, the
dash-dotted curves are the TU and TL formulas, and the solid curve
is the new CG-GDF formula that we will describe below.

Regarding SSFM simulations, we first note a shift of the CG SNR with
respect to the COP SNR, much in line with the results in \cite[Fig 2(b)]{ASN_droop_ofc19}.
The low-power SNR of the CG case is larger than the COP case because
it does not experience ASE-induced droop; the high-power SNR of the
CG case is smaller than the COP case because of the span-by-span increase
of power in CG that generates nonlinearity.

Regarding the T-formulas, we note that TU overestimates the simulated
CG SNR, while TL with depletion under-estimates the CG SNR. All the
CG curves (simulations, T-formulas, CG-GDF) at low powers tend to
coincide with the theoretical GN curve, as they should.

\subsection{\label{subsec:The-CG-GDF}The CG-GDF and the improved TU}

The trouble with the T-formulas is that they try to model a strongly
nonlinear system with an amended end-to-end RP1 system. The amendments,
however, do contain the correct physical intuition. So the key to
the new CG-GDF is to use the intuition \cite{poggio_depletion-2}
about the effective power for NLI generation, eq. (\ref{eq:PNLI_carena}),
but with a re-normalized RP1 per-span model instead of the end-to-end
RP1 GN model.

In CG mode we can consider only the ASE and NLI on the per-tributary
bandwidth (O-ASE does not affect the SNR). The power-flow diagram
is again given by Fig. \ref{fig:block}(d), where now we have $\chi_{a}=1$,
i.e., no ASE droop, and $\delta P_{i}\mathcal{L}^{-1}\equiv\beta$.
As in (\ref{eq:PNLI_carena}), we now let $\delta P_{rk}=\alpha_{NL}(P_{t}+(k-1)\beta)^{3}$.

The new SNR formula, which we call the \emph{CG-GDF,} is thus calculated
by eq.\foreignlanguage{english}{ (\ref{eq:eGDF}), where we now use
$\chi_{k}=\chi_{rk}$, $\eta_{A}=1$, and replace 
\begin{equation}
\left\{ \begin{array}{l}
(\chi_{a}^{-1}-1)\to\beta/P_{t}\\
\chi_{rk}^{-1}\to1+\alpha_{NL}P_{t}^{2}(1+(k-1)\frac{\beta}{P_{t}})^{3}.
\end{array}\right.\label{eq:chi_CG-GDF}
\end{equation}
}

\selectlanguage{english}%
In the example of case A), \foreignlanguage{american}{Fig. \ref{fig:CG_finale}
shows an excellent match between CG-simulations and the CG-GDF. A
similarly good match is obtained in the remaining cases B and C.}

\section{\label{sec:Conclusions}Conclusions}

In this paper, we presented an analytical model to fully theoretically
support the GDF disclosed by Antona \emph{et al.} in \cite{ASN_droop_ofc19,ASN_suboptic},
which includes various fiber power-redistribution mechanisms such
as Kerr nonlinearity, GAWBS, and internal and external crosstalk.
We verified the GDF against simulations in three published case studies
drawn from \cite{ASN_droop_ofc19,amir_droop}.

We provided upper and lower bounds to the GDF. Using the tightest
upper-bound we provided analytical expressions of the SNR and Shannon
spectral efficiency (SE) gaps from the GN formula to the GDF. We showed
that the gaps can be effectively expressed only in terms of the GN-predicted
SNR and SE. We showed that the SE gap (per mode) is bounded between
0.6 and 0.9 b/s/Hz when the GN-SE is as low as 2 b/s/Hz, and it decreases
as the GN-SE increases, e.g., it is less than 0.2 b/s/Hz when the
GN-SE is above 6 b/s/Hz.

We extended the GDF to the case where ASE has larger bandwidth/mode
occupancy than the signal. The resulting COP-GDF equation depends
only on the amplifier fill-in efficiency $\eta_{A}$, eq. (\ref{eq:etaA}),
and was found to very well match simulations in all considered cases.
Finally, we extended the theory to include constant-gain amplifier
chains, and we derived the new CG-GDF SNR expression that matches
simulations better than any other previously known formula.

One of the key theoretical results is that the end-to-end model underlying
all GDF expressions is a concatenation of per-span RP1 models with
end-span power renormalization. This fact allows the GDFs to well
reproduce the SNR of highly nonlinear systems, well beyond the RP1
limit of the GN model. The model is reminiscent of multi-stage backpropagation,
that combines the benefits of split-step and perturbation-based approaches.

\section*{Appendix A: GDF for inhomogeneous spans}

\selectlanguage{american}%
The generalization of the GDF model to the inhomogeneous case goes
as follows. The $k$-th span now has input power $P_{k-1}$, fiber
span loss $\mathcal{L}{}_{k}<1$ and an end-span amplifier operated
in COP mode with fixed output power $P_{k}$, $k=0,...,N$ (where
$P_{0}$ is the launched power) and a gain $G_{k}$ which, in absence
of ASE-induced droop, would be $G_{k}=P_{k}/(P_{k-1}\mathcal{L}_{k})$.
With ASE droop the gain is smaller by a factor $\chi_{ak}<1$. The
power flow diagram for the inhomogeneous case is similar to the homogenous
case of Fig. \foreignlanguage{english}{\ref{fig:block}, where i)
all quantities are span-$k$ dependent; ii) the multiplicative output
factor in diagrams (c), (d) is now $\chi_{ak}P_{k}/P_{k-1}$ instead
of only $\chi_{a}$; iii) in diagram (d) the input sub-block has input/output
power $P_{k-1}$, while the output sub-block has $P_{k-1}$ in and
$P_{k}$ out.}

From the modified diagram (d) we thus derive:

1) the power balance at the output sub-block: $(P_{k-1}+\delta P_{ik}\mathcal{L}_{k}^{-1})\chi_{ak}\frac{P_{k}}{P_{k-1}}=P_{k}$,
which yields
\begin{equation}
\chi_{ak}=(1+\frac{\delta P_{ik}\mathcal{L}_{k}^{-1}}{P_{k-1}})^{-1}\triangleq(1+SNR_{a1k}^{-1})^{-1}\label{eq:chia-1}
\end{equation}
where we implicitly defined the SNR degraded by ASE at the single
amplifier as 
\begin{equation}
SNR_{a1k}\triangleq\frac{P_{k-1}}{\delta P_{ik}\mathcal{L}_{k}^{-1}}.\label{eq:SNRa1-1}
\end{equation}

2) the power balance at the input sub-block: $(P_{k-1}+\delta P_{rk})\chi_{rk}=P_{k-1}$,
which yields
\begin{equation}
\chi_{rk}=(1+\frac{\delta P_{rk}}{P_{k-1}})^{-1}\triangleq(1+SNR_{r1k}^{-1})^{-1}\label{eq:chir-1}
\end{equation}
where we implicitly defined the SNR degraded by power redistribution
at the single amplifier as 
\begin{equation}
SNR_{r1k}\triangleq\frac{P_{k-1}}{\delta P_{rk}}.\label{eq:SNRr1-1}
\end{equation}

Define the span total droop as the product of addition and redistribution
droops: $\chi_{k}\triangleq\chi_{rk}\chi_{ak}.$ The power block diagram
of the $k-$th span in (modified) diagram (d) shows that the total
span power-gain seen by the transiting signal is $(P_{k}/P_{k-1})\chi_{k}$,
hence the desired multiplex signal power at the output of the $N$-th
amplifier is 
\[
P_{s}(N)=P_{0}\prod_{k=1}^{N}(P_{k}/P_{k-1})\chi_{k}=P_{N}\prod_{k=1}^{N}\chi_{k}
\]
so that the total noise power after $N$ spans is $P_{a}(N)+P_{r}(N)=P_{N}(1-\prod_{k=1}^{N}\chi_{k})$.

Hence the\emph{ }OSNR at the output of the chain from amplifiers 1
to $N$, i.e., the ratio of total multiplex signal power to total
noise power at the output of the $N$-th amplifier, is obtained as
\begin{equation}
OSNR=\frac{1}{\left[\prod_{k=1}^{N}(1+\frac{1}{SNR_{a1k}})(1+\frac{1}{SNR_{r1k}})\right]-1}\label{eq:GDF-2}
\end{equation}
which leads to the general product rule for inverse droops, eq. (\ref{eq:PRID-1})
in the main text.

For the calculation of the per-tributary SNR it is necessary instead
to have the individual expression of $P_{a}(N)$, $P_{r}(N)$, as
seen in eq. (\ref{eq:tributarySNR}) in the main text. It is possible
to read off the modified diagram (d) the update rule for useful signal,
additive and redistribution noise at any span $k$ as:
\begin{align}
P_{s}(k) & =P_{s}(k-1)\chi_{k}(P_{k}/P_{k-1})\nonumber \\
P_{a}(k) & =(P_{a}(k-1)+\delta P_{ik}\mathcal{L}_{k}^{-1}\chi_{rk}^{-1})\chi_{k}(P_{k}/P_{k-1})\label{eq:recursions}\\
P_{r}(k) & =(P_{r}(k-1)+\delta P_{rk})\chi_{k}(P_{k}/P_{k-1})\nonumber 
\end{align}
with initial conditions: $P_{s}(0)=P_{0}$, $P_{a}(0)=P_{r}(0)=0$.
\foreignlanguage{english}{The second and third recursions are of the
kind: $u(k)=(u(k-1)+b_{k})a_{k}$, whose general solution when $u(0)=0$
is: $u(N)=\sum_{k=1}^{N}b_{k}\prod_{m=k}^{N}a_{m}$.}

\selectlanguage{english}%
Hence, using $\delta P_{ik}\mathcal{L}_{k}^{-1}\equiv P_{k-1}(\chi_{ak}^{-1}-1)$
and $\delta P_{rk}\equiv P_{k-1}(\chi_{rk}^{-1}-1)$ from (\ref{eq:chia-1}),(\ref{eq:chir-1}),
the formal solutions are 
\begin{align}
P_{s}(N) & =P_{N}\prod_{m=1}^{N}\chi_{m}\nonumber \\
P_{a}(N) & =P_{N}\sum_{k=1}^{N}(\chi_{ak}^{-1}-1)\chi_{rk}^{-1}\prod_{m=k}^{N}\chi_{m}\label{eq:SOLrecursions}\\
P_{r}(N) & =P_{N}\sum_{k=1}^{N}(\chi_{rk}^{-1}-1)\prod_{m=k}^{N}\chi_{m}\nonumber 
\end{align}

\selectlanguage{american}%
Although this closed-form solution is pleasing, the recursion (\ref{eq:recursions})
is the one we use for calculations.

\section*{Appendix B: Caveats in $\alpha_{NL}$ estimation from SSFM simulations}

This appendix discusses some caveats in estimating from SSFM simulations
the span-average nonlinear coefficient $\alpha_{NL}$ to be used in
the theoretical formulas, whenever the end-to-end system is highly
nonlinear, way beyond the RP1 limits. Let the generic tributary complex
received field be $A_{rx}(t)$. In absence of ASE we know from the
GDF model that we have only redistribution droop: $\chi\equiv\chi_{r}=(1+\alpha_{NL}P^{2})^{-1}$,
and the received power is
\begin{align*}
<|A_{rx}(t)|^{2}> & =P=\chi^{N}P+(1-\chi^{N})P\\
 & \equiv\chi^{N}P(1+V_{NLI})
\end{align*}
where the brackets denote time averaging, and we implicitly defined
the variance of the normalized end-to-end NLI 
\begin{equation}
V_{NLI}(P)\triangleq\chi^{-N}-1=(1+\alpha_{NL}P^{2})^{N}-1.\label{eq:eq_vnli}
\end{equation}

Now define\footnote{Please note the use of boldface, to distinguish $\boldsymbol{\alpha}_{NL}(P)$
which is $P$ dependent, from the span-average coefficient $\alpha_{NL}$
which is power independent.} 
\begin{equation}
\boldsymbol{\alpha}_{NL}(P)\triangleq\frac{V_{NLI}(P)}{NP^{2}}=\frac{(1+\alpha_{NL}P^{2})^{N}-1}{NP^{2}}\label{eq:expression}
\end{equation}
as the\emph{ span-averaged power-dependent NLI coefficient}. At small
powers it coincides with the NLI coefficient $\alpha_{NL}$ of the
\emph{per-span} RP1 expansion, since $\lim_{P\to0}V_{NLI}(P)/(NP^{2})=\alpha_{NL}$.
In a truly end-to-end RP1 system, the quantity $\boldsymbol{\alpha}_{NL}(P)$
should at all powers coincide with $\alpha_{NL}$. At powers for which
$\boldsymbol{\alpha}_{NL}(P)$ markedly departs from $\alpha_{NL}$,
the end-to-end line ceases to be RP1.

An estimation $\hat{V}_{NLI}(P)$ of the value of $V_{NLI}(P)$ in
(\ref{eq:expression}) is routinely obtained from the samples of the
received constellation scatter diagram.

\selectlanguage{english}%
\begin{figure}
\centering

\includegraphics[width=1\columnwidth]{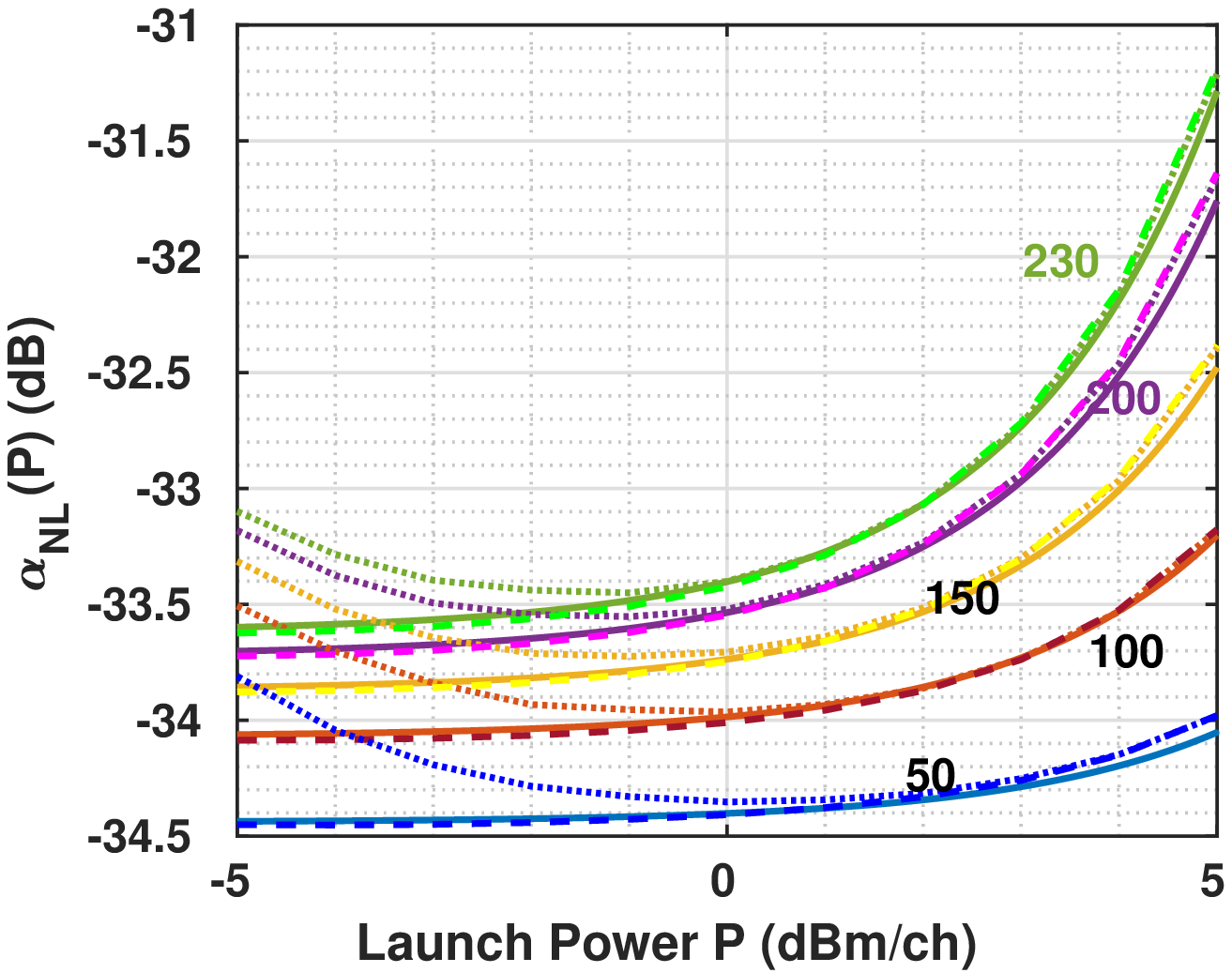}

\caption{\label{fig:sim_aNL} \foreignlanguage{american}{dB-value of span-averaged
power-dependent NLI coefficient $\boldsymbol{\alpha}_{NL}(P)$ }{[}mW$^{-2}${]}
in eq. (\ref{eq:expression}) versus launch power $P$ (thick solid)\foreignlanguage{american}{
and its estimate $\hat{\boldsymbol{\alpha}}_{NL}(P)\triangleq\hat{V}_{NLI}(P)/(NP^{2})$
}from simulated scatter diagrams in absence of ASE\foreignlanguage{american}{
(dotted for a step-size with nonlinear phase $10^{-3}$(rad), dashed
for $5\cdot10^{-4}$(rad))}. \foreignlanguage{american}{Values used
in theoretical formula: $\alpha_{NL}=${[}-34.44, -34.07, -33.87,
-33.72, -33.62{]} }(dB) for $N=${[}50, 100, 150, 200, 230{]} spans.}
\end{figure}

\selectlanguage{american}%
We ran SSFM simulations of the PDM-QPSK system in \cite{ASN_droop_ofc19},
case A. Fig. \ref{fig:sim_aNL} shows a plot of the estimated span-averaged
power-dependent NLI coefficient $\hat{\boldsymbol{\alpha}}_{NL}(P)\triangleq\hat{V}_{NLI}(P)/(NP^{2})$
\foreignlanguage{english}{{[}mW$^{-2}${]}} (dotted for a standard
step-size with nonlinear phase per span (NLP) $10^{-3}$(rad), dashed
for $5\cdot10^{-4}$(rad)) and the theoretical expression (\ref{eq:expression})
in thick solid line, in which the best-fitting values $\alpha_{NL}=${[}-34.44,
-34.07, -33.87, -33.72, -33.62{]} \foreignlanguage{english}{(dB)}
for 5 values of the span number $N=[50,100,150,200,230]$ were used
to match the low-power values of $\boldsymbol{\alpha}_{NL}(P)$ to
simulations with the most precise NLP. Such best-fitting $\alpha_{NL}$
values are those used in the theoretical formulas in the text, and
are reported in Fig. \ref{fig:sim_aNL_vs_N} along with those of cases
B) and C).

Fig. \ref{fig:sim_aNL} highlights two interesting facts:

1) for the system in case A, with a standard choice of the step size
the estimated variance does not flatten out at low powers, as it should
when the end-to-end line behaves as a truly RP1 system. So care should
be taken to verify that the simulation-based estimation of $\alpha_{NL}$
is done with the correct step size;

2) when the end-to-end system is more nonlinear than an RP1 system,
the new analytical expression (\ref{eq:expression}) for $\boldsymbol{\alpha}_{NL}(P)$
well matches the simulated quantity $\hat{\boldsymbol{\alpha}}_{NL}(P)\triangleq\hat{V}_{NLI}(P)/(NP^{2})$,
a further indication that the locally-RP1, power-renormalized concatenation
underlying the GDF well models such a highly nonlinear system.

In practice, normally only a single low-power end-to-end simulation
(with a correct step-size) can be run to estimate the value of the
span-averaged $\alpha_{NL}$ to be used in the theory as $(\hat{V}_{NLI}(P)+1)^{1/N}-1)/P^{2}$,
as per (\ref{eq:eq_vnli}). Alternatively, one may use the EGN model
\foreignlanguage{english}{\cite{EGN1_torino,EGN2_torino,EGN_dar,EGN_serena}}
to get the end-to-end NLI coefficient and then divide the result by
$N$ to get the span-average $\alpha_{NL}$.
\selectlanguage{english}%

\section*{Appendix C: SE gap approximations}

\selectlanguage{american}%
This appendix derives two approximations of the SE decrease due to
droop with respect to the GN value, as shown in Sec. \ref{subsec:Comparisons-with-the}.

The SE decrease (at any power $P$, not only at top value $P_{0}$)
is:
\[
\Delta SE\equiv SE_{GN}-SE_{GDF}=2\log_{2}(\frac{1+SNR_{GN}}{1+SNR_{GDF}})
\]
Thus using the UB (\ref{eq:snr_UB}) and  at large $N$ we get:
\begin{align*}
\Delta SE & \geq2\log_{2}(\frac{1+SNR_{GN}}{1+\frac{2\left(SNR_{GN}\right)^{2}}{1+2SNR_{GN}}})\\
 & =2\log_{2}(\frac{1+3SNR_{GN}+2\left(SNR_{GN}\right)^{2}}{1+2SNR_{GN}+2\left(SNR_{GN}\right)^{2}})\\
 & =\frac{2}{\ln(2)}\ln(1+\frac{SNR_{GN}}{1+2SNR_{GN}+2\left(SNR_{GN}\right)^{2}})\\
 & \leq\frac{2}{\ln(2)}\frac{SNR_{GN}}{1+2SNR_{GN}+2\left(SNR_{GN}\right)^{2}}
\end{align*}
\foreignlanguage{english}{which is approximation (\ref{eq:DSE}) in
the main text. It is neither an upper nor a lower bound.}

\selectlanguage{english}%
Using instead the LB (\ref{eq:SNR_LB}) at large $N$ we get
\begin{align*}
\Delta SE & \leq2\log_{2}(\frac{1+SNR_{GN}}{1+SNR_{GN}-\frac{1}{2}})\\
 & =2\log_{2}(1+\frac{1/2}{SNR_{GN}+\frac{1}{2}})\\
 & \leq\frac{1}{\ln(2)[SNR_{GN}+\frac{1}{2}]}
\end{align*}
which is the UB (\ref{eq:DSE_UB}).
\selectlanguage{american}%


\begin{thebibliography}{10}
\bibitem{giles}C. R. Giles and E. Desurvire, ``Propagation of Signal
and Noise in Concatenated Erbium-Doped Fiber Optical Amplifiers'',
\emph{J. Lightw. Technol.}, vol. 9, no. 2, pp. 147-154, Feb. 1991.

\bibitem{amir_droop}A. Ghazisaeidi, ``A Theory of Nonlinear Interactions
between Signal and Amplified Spontaneous Emission Noise in Coherent
Wavelength Division Multiplexed Systems,'' \emph{J. Lightw. Technol.},
vol. 35, pp. 5150--5175, Dec. 2017.

\bibitem{TEsubcom_ptl17}O. V. Sinkin \emph{et al.}, ``Maximum Optical
Power Efficiency in SDM-Based Optical Communication Systems,'' \emph{Photon.
Technol. Lett.}, vol. 29, no. 13, pp. 1075-1077, Jul. 2018.

\bibitem{TEsubcom_jlt18}O. V. Sinkin \emph{et al.}, ``SDM for Power-Efficient
Undersea Transmission,'' \emph{J. Lightw. Technol.}, vol. 36, no.
2, pp. 361-371, Jan. 2018.

\bibitem{ASN_droop_ofc19}J.-C. Antona, A. Carbo Méseguer, and V.
Letellier, \textquotedbl Transmission Systems with Constant Output
Power Amplifiers at Low SNR Values: a Generalized Droop Model,\textquotedbl{}
in\foreignlanguage{english}{ Proc. \emph{Opt. Fiber Commun. (OFC)},
San Diego (CA), 2019, paper }M1J.6\foreignlanguage{english}{.}

\bibitem{ASN_suboptic}J.-C. Antona et al., Performance of open cable:
from modeling to wide scale experimental assessment,'' in \emph{Proc.
SubOptic}, New Orleans (LA), 2019.

\bibitem{poggio_ptl}G. Bosco \emph{et al}., ``Performance prediction
for WDM PM-QPSK transmission over uncompensated links,'' in\foreignlanguage{english}{
Proc. \emph{Opt. Fiber Commun. (OFC)}, San Diego (CA), 2011, paper
OThO7.}

\bibitem{grellier}E. Grellier and A. Bononi, ``Quality parameter
for coherent transmissions with Gaussian-distributed nonlinear noise,''\emph{
Opt. Exp}., vol. 19, no. 13, pp. 12781-{}-12788, Jun. 2011.

\bibitem{GN_carena}A. Carena, \emph{et al.}, \textquotedblleft Modeling
of the Impact of Non-Linear Propagation Effects in Uncompensated Optical
Coherent Transmission Links,\textquotedblright{} \emph{J. Lightw.
Technol}., vol. 30, no. 10, pp. 1524-1539, may 2012.

\bibitem{GN_poggio}P. Poggiolini, \textquotedblleft The GN model
of non-linear propagation in uncompensated coherent optical systems,\textquotedblright{}
\foreignlanguage{english}{\emph{J. Lightw. Technol}., vol. 30, no.
24, pp. 3857--3879, Dec. 2012}.

\bibitem{noiECOC20_subm}A. Bononi, J.-C. Antona, A. Carbo Méseguer,
P. Serena, ``A model for the generalized droop formula,'' in Proc.
\foreignlanguage{english}{\emph{European Conf. on Optical Commun.}
\emph{(ECOC)}, Dublin, Ireland, 2019, paper W.1.D.5. Also available
at }arXiv:1906.08645.

\bibitem{noi_RP}A. Vannucci, P. Serena, and A. Bononi, ``The RP
method: a new tool for the iterative solution of the nonlinear Schroedinger
equation,``\emph{ J. Lightw. Technol.}, vol. 20, pp. 1102--1112,
Jul. 2002.

\bibitem{poggio_arxiv}P. Poggiolini, G. Bosco, A. Carena, V. Curri,
F. Forghieri, \textquotedblleft A Detailed Analytical Derivation of
the GN Model of Non-Linear Interference in Coherent Optical Transmission
Systems,\textquotedblright{} Available: arXiv:1209.0394 , {[}physics.optics{]}
(2012).

\bibitem{johannisson_GN}P. Johannisson and M. Karlsson, \textquotedblleft Perturbation
analysis of nonlinear propagation in a strongly dispersive optical
communication system,\textquotedblright{} \emph{J. Lightw. Technol.}
vol. 31, pp. 1273-1282 (2013).

\bibitem{RP1_GN_arxiv}A. Bononi and P. Serena, \textquotedblleft An
alternative derivation of Johannisson\textquoteright s regular perturbation
model,\textquotedblright{} Available: arXiv:1207.4729, {[}physics.optics{]}
(2012).

\bibitem{secondini}M. Secondini, D. Marsella, and E. Forestieri,
\textquotedblleft Enhanced split-step Fourier method for digital backpropagation,\textquotedblright{}
\foreignlanguage{english}{in Proc. \emph{European Conf. on Optical
Commun.} \emph{(ECOC)}, Cannes, France, 2014, paper} We.3.3.5.

\bibitem{kumar} X. Liang and S. Kumar, \textquotedblleft Multistage
perturbation theory for compensating intra-channel impairments in
fiber optic systems,\textquotedblright{} \emph{Opt. Exp.}, vol. 22,
no. 24, pp. 29733-29745, Dec. 2014.

\selectlanguage{english}%
\bibitem{poggio_depletion-2}P. Poggiolini, A. Carena, Y. Jiang, G.
Bosco, V. Curri, and F. Forghieri, \textquotedblleft Impact of low-OSNR
operation on the performance of advanced coherent optical transmission
systems,'' in Proc. \emph{European Conf. on Optical Commun.} \emph{(ECOC)},
Cannes, France, 2014, paper Mo.4.3.2.

\selectlanguage{american}%
\bibitem{serena_S-NLI_interaction}P. Serena, \textquotedblleft Nonlinear
Signal--Noise Interaction in Optical Links With Nonlinear Equalization
,'' \emph{J. Lightw. Technol}., vol. 34, no. 6, pp. 1476-{}-1483,
Mar. 2016.

\bibitem{nyq-WDM}G. Bosco, V. Curri, A. Carena, P. Poggolini, and
F. Forghieri, ``On the Performance of Nyquist-WDM Terabit Superchannels
Based on PM-BPSK, PM-QPSK, PM-8QAM or PM-16QAM Subcarriers,'' \emph{J.
Lightw. Technol.}, vol. 29, n. 1, pp. 53-61 (2011).

\bibitem{Haus}H. A. Haus, ``The Noise Figure of Optical Amplifiers.''
\emph{Photon. Technol. Lett.}, vol. 10, no. 11, pp. 1602-1604, Nov.
1998.

\bibitem{Vacondio}F. Vacondio \emph{et al.}, \textquotedbl On nonlinear
distortions of highly dispersive optical coherent systems,'' \emph{Opt.
Exp}., vol. 20, no. 2, pp. 1022-{}-1032, Jan. 2012.

\bibitem{GAWBS}M. A. Bolshtyansky \emph{et al}., ``Impact of Spontaneous
Guided Acoustic-Wave Brillouin Scattering on Long-haul Transmission,``
\foreignlanguage{english}{in Proc. \emph{Opt. Fiber Commun. (OFC)},
San Diego (CA), 2018, paper }M4B.3\foreignlanguage{english}{.}

\bibitem{MMF_Xtalk}J. M. Gené and P. J. Winzer ``A Universal Specification
for Multicore Fiber Crosstalk,'' \emph{Photon. Technol. Lett.}, vol
31, n. 9, pp. 673-676 (2019).

\bibitem{EGN1_torino}A. Carena \emph{et al.} ``EGN model of non-linear
fiber propagation,'' \emph{Opt. Exp}., vol. 22, no. 13, pp. 16335-{}-16362,
Jun. 2014.

\bibitem{EGN2_torino}P. Poggiolini \emph{et al.,} ``A Simple and
Effective Closed-Form GN Model Correction Formula Accounting for Signal
Non-Gaussian Distribution,'' \emph{J. Lightw. Technol}., vol. 33,
no. 2, pp. 459-{}-473, Jan. 2015.

\bibitem{EGN_dar}R. Dar, M. Feder, A. Mecozzi, and M. Shtaif, ``Accumulation
of nonlinear interference noise in fiber-optic systems,'' \emph{Opt.
Exp}., vol. 22, no. 12, pp. 14199-{}-14211, Jun. 2014.

\bibitem{EGN_serena}P. Serena, and A. Bononi, ``A Time-Domain Extended
Gaussian Noise Model,'' \emph{ J. Lightw. Technol.}, vol. 33, no.
7, pp. 1459-{}-1472, Apr. 2015.
\end{thebibliography}
\end{document}